\documentclass[aps,pra,twocolumn,superscriptaddress]{revtex4}
\usepackage{amsfonts,amssymb,graphicx}
\usepackage[usenames,dvipsnames]{xcolor}
\usepackage{pdfpages}
\usepackage{footmisc}
\usepackage[normalem]{ulem}
\usepackage{color}

\begin{document}

\title{Exact one- and two-site reduced dynamics in a finite-size quantum Ising ring after a quench: A semi-analytical approach}
\author{Ning Wu}
\email{wunwyz@gmail.com}
\affiliation{Center for Quantum Technology Research, School of Physics, Beijing Institute of Technology, Beijing 100081, China and Key Laboratory of Advanced Optoelectronic Quantum Architecture and Measurements (MOE), School of Physics, Beijing Institute of Technology, Beijing 100081, China}
\begin{abstract}
We study the non-equilibrium dynamics of a homogeneous quantum Ising ring after a quench, in which the transverse field $g$ suddenly changes from zero to a nonzero value. The long-timescale reduced dynamics of a single spin and of two nearest-neighbor spins, which involves the evaluation of expectation values of odd operators that break the fermion parity, is exactly obtained for finite-size but large rings through the use of a recently developed Pfaffian method [N. Wu, Phys. Rev. E \textbf{101}, 042108 (2020)]. Time dependence of the transverse and longitudinal magnetizations ($\langle\sigma^z_j\rangle_t$ and $\langle\sigma^{x,y}_j\rangle_t$), single-spin purity, expectation value of the string operator $X_j=\prod^{j-1}_{l=1}\sigma^z_l\sigma^x_j$ ($\langle X_j\rangle_t$), several equal-time two-site correlators ($\langle\sigma^{x,z}_j\sigma^{x,z}_{j+1}\rangle_t$, $\langle\sigma^x_j\sigma^y_{j+1}\rangle_t$, and $\langle\sigma^x_j\sigma^z_{j+1}\rangle_t$), and pairwise concurrence after quenches to different phases are numerically studied. Our main findings are that (i) The expectation value of a generic odd operator approaches zero in the long-time limit; (ii) $\langle X_j\rangle_t$ exhibits $j$-independent exponential decay for a quench to $g=1$ and the time at which $\langle X_j\rangle_t$ reaches its first maximum scales linearly with $j$; (iii) The single-spin purity dynamics is mainly controlled by $\langle\sigma^x_j\rangle_t$ ($\langle\sigma^z_j\rangle_t$) for a quench to $g<1$ ($g\geq 1$). For quenches to the disordered phase with $g\gg1$, the single-spin tends to be in the maximally mixed state and the transverse and longitudinal correlators $\langle\sigma^z_j\sigma^z_{j+1}\rangle_t$ and $\langle\sigma^x_j\sigma^x_{j+1}\rangle_t$ respectively approaches $-0.25$ and $0.5$ in the thermodynamic limit; (iv) The nearest-neighbor entanglement acquires a finite plateau value that increases with increasing $g$, and approaches a saturated value $\sim0.125$ for $g\gg1$.
\end{abstract}

\maketitle
\section{Introduction}
\par The non-equilibrium dynamics of isolated quantum many-body systems is a subject under intense theoretical and experimental study in the past decades. The experimental advances in cold atom systems, nanoscience, and quantum optics enable possible realizations of quantum many-body models in these platforms~\cite{RMP2011}. Exactly solvable models serve as ideal testbeds for the theoretical investigation of non-equilibrium protocols. The one-dimensional quantum Ising model (or more generally the quantum $XY$ chain) is perhaps the simplest soluble model exhibiting a quantum phase transition~\cite{QPTbook} and offers a suitable platform for investigating a variety of non-equilibrium quantum phenomena, including the Kibble-Zurek mechanism~\cite{KZM1,KZM2,Damski2020}, adiabatic transitions~\cite{Trans1,Trans2,Trans3}, quantum quenches~\cite{quench1,JSM2012,quench3,Rossini2020}, dynamical quantum phase transitions~\cite{DQPT1,DQPT2}, quantum chaos~\cite{Lin2018}, and discrete time crystals~\cite{DTC1,Yu2019}, etc.
\par A quantum Ising ring can be mapped to a spinless fermion model via the Jordan-Wigner (JW) transformation~\cite{Lieb1961}. However, the resulting fermion model does not admit a simple cyclic structure due to the presence of the boundary term. Instead, one gets two fermion chains with periodic and antiperiodic boundary conditions, which respectively support even and odd numbers of fermions. A commonly used procedure for evaluating the real-time dynamics in large spin rings is to separately calculate the dynamics of each uncoupled momentum mode of the fermion chains. However, due to the nonlocal nature of the JW transformation, the relationship between states or observables in the spin representation and those in the momentum space is often highly complicated. Even if the spin state of interest is expressible in terms of the momentum-space eigenstates, one still needs to convert the observables (usually local spin operators) into the momentum representation, which is often a difficult task.
\par Besides the above-mentioned issues, additional difficulty arises when one tries to evaluate expectation values of odd operators that break the fermion parity in a state having both even- and odd-parity components. This issue may be dated back to the 1970's when McCoy \emph{et al}. tried to analytically calculate the time-dependent longitudinal correlation function in a finite-size $XY$ chain~\cite{PRA1971}. Instead of directly attacking the problem of calculating matrix elements of odd operators, the authors of Ref.~\cite{PRA1971} used a so-called `doubling trick' to consider the four-spin correlation in the thermodynamic limit, which involves only even operators, and hence can be treated by using standard free-fermion techniques~\cite{Lieb1961}. The same trick was recently employed to study the longitudinal out-of-time-ordered correlators in a quantum Ising ring~\cite{Lin2018}.
\par The dynamics of longitudinal magnetization (an odd operator) starting with $Z_2$ symmetry-breaking initial states (superpositions of even and odd states) in the quantum Ising ring has recently attracted much attention in various contexts~\cite{Damski2020,quench1,JSM2012,Rossini2020,DQPT1,Yu2019,Eisler2018}. Because of the invalidity of free-fermion techniques mentioned above, most of these works either use exact diagonalization to track the time evolution in small rings~\cite{Damski2020,Rossini2020,Yu2019}, or employ advanced analytical techniques to obtain the dynamics in the thermodynamic limit~\cite{JSM2012,Eisler2018}. With the intention of efficiently calculating the longitudinal magnetization dynamics in finite-size but large quantum Ising rings, the author recently found that the matrix element of the longitudinal magnetization between states with distinct fermion parities can be expressed as the Pfaffian of an appropriate matrix whose entries can be analytically obtained~\cite{PRE2020}. This provides an efficient method to calculate the long-timescale longitudinal magnetization dynamics in large rings far beyond the reach of exact diagonalization.
\par In this work, we study the dynamics of a quantum Ising ring after a sudden quench of the transverse field from zero to a finite value $g$. Sudden quench is the simplest nontrivial protocol for inducing non-equilibrium dynamics in isolated many-body systems and has been widely studied in the literature. It is closely related to a variety of important physical phenomena, such as thermalization and relaxation towards a steady state in integrable quantum spin systems~\cite{Adv2010,PRL2011,JSM2016}, entanglement dynamics after a quantum quench~\cite{Alba2014,Coser2014,JSM2017,Alba2018}, and the interplay between the two~\cite{Science2016,PNAS2017}. The feasibility of performing numerical simulations of quench dynamics in sufficiently large systems and over long time scales is essential to the understanding of these phenomena.
\par To be specific, we choose the initial state as one of the two degenerate ferromagnetic ground states of the classical Ising ring, which breaks the $Z_2$ symmetry and thus contains both even and odd component states in the fermion representation. We obtain exact dynamics of reduced density matrices of both a single spin and of two nearest-neighbor spins. The time evolution of the single-spin reduced density matrix is simply determined by the polarization dynamics $\langle\vec{\sigma}_j\rangle_t$. The longitudinal magnetizations $\langle\sigma^{x,y}_j\rangle_t$ are calculated using the Pfaffian method~\cite{PRE2020}. More generally, we obtain the quenched dynamics of the string operator $X_j=\prod^{j-1}_{l=1}\sigma^z_l\sigma^x_j$ (an odd operator), which involves a subsystem of length $j$, but is still a local operator~\cite{JSM2016}. It is found that $\langle X_j\rangle_t$ exhibits $j$-independent exponential decay in the long time limit for a quench to the critical point $g=1$. Due to the finite-size nature of our simulations, we observe the collapse of $\langle X_j\rangle_t$ within the time interval $t\in[j/2,(N-j)/2]$ ($N$ is the total number of sites) and its partial revival around the time $t=N/2$. These features are consistent with Cardy's general analysis based on conformal field theory~\cite{Cardy2014}. For $g=1$, we also find that the time at which $\langle X_j\rangle_t$ reaches its first maximum scales linearly with the string length $j$, while these maxima decay exponentially with increasing $j$.
\par Analysis on the single-spin purity dynamics $P(t)$ shows that the longitudinal magnetization $\langle\sigma^x_j\rangle_t$ (transverse magnetization $\langle\sigma^z_j\rangle_t$) mainly controls the overall profile of $P(t)$ for quenches with $g<1$ ($g\geq 1$). For quenches to the deeply disordered phase with $g\gg1$, the single-spin rapidly approaches a maximally mixed state with nearly vanishing polarization. Important and interesting features in the long-time dynamics arise for quenches into the vicinity of the critical point in large enough systems. Based on a determinant approach, Calabrese \emph{et. al.} show that for quenches within the ordered phase $\langle\sigma^x_j\rangle_t$ relaxes to zero exponentially at long times and derive explicit expressions for the prefactor and decay function in the thermodynamics limit~\cite{JSM2012}. Recently, Rossini and Vicari revisited the same problem using exact diagonalization on a ring with $N=24$ sites~\cite{Rossini2020}. They show that both the prefactor and the decay function show singular features at $g=1$. Our numerical simulations on much larger systems indeed confirm the existence of the discontinuity discovered in Ref.~\cite{Rossini2020}.
\par Determination of the reduced density matrix of two nearest-neighbor spins is helpful in studying the relationship between entanglement and quantum critical phenomena~\cite{PRA2002,Nature2002,PRA2010}. Although the diagonal elements of the two-spin reduced density matrix can be calculated using free-fermion techniques, some of the off-diagonal elements actually involve products of three fermion operators, which makes the evaluation of these matrix elements seemingly formidable. Nevertheless, the translational and spatial inversion invariance of both the Hamiltonian and initial state allows us to express expectation values of the `triple' fermionic operators in terms of those of `single' ones so that the Pfaffian technique is still applicable. Based on the obtained two-spin reduced density matrix, we study the time dependence of various equal-time two-spin correlators after the quench, including the transverse and longitudinal correlators $\langle\sigma^z_j\sigma^z_{j+1}\rangle_t$ and $\langle\sigma^x_j\sigma^x_{j+1}\rangle_t$, and the cross correlators $\langle\sigma^x_j\sigma^y_{j+1}\rangle_t$ and $\langle\sigma^x_j\sigma^z_{j+1}\rangle_t$. We derive an analytical expression for $\langle\sigma^z_j\sigma^z_{j+1}\rangle_t$, which in the thermodynamic limit can be written as a double integral over the momentum variables. Under general quench protocols both $\langle\sigma^z_j\sigma^z_{j+1}\rangle_t$ and $\langle\sigma^x_j\sigma^x_{j+1}\rangle_t$ approach nonzero plateau values in large enough rings, in spite of the accompanying disappearance of $\langle\sigma^z_j \rangle_t$ and $\langle\sigma^x_j \rangle_t$. In addition, the steady values of $\langle\sigma^z_j\sigma^z_{j+1}\rangle_t$ and $\langle\sigma^x_j\sigma^x_{j+1}\rangle_t$ respectively saturate to $-0.25$ and $0.5$ for quenches to large enough $g$, which is quantitatively explained by analyzing the corresponding analytical expressions in the thermodynamic and long-time limits. We finally study the nearest-neighbor entanglement dynamics after the quench and find that the entanglement is steadily generated after quenches to the disordered phase.
\par The rest of the paper is organized as follows. In Sec.~\ref{SecII}, we introduce the quantum Ising model and briefly review its diagonalization. We also introduce our initial state and obtain the analytical expression for the time evolved state in the momentum space. In Sec.~\ref{SecIII}, we study the single-spin reduced dynamics in detail based on the obtained longitudinal and transverse magnetizations. The quench dynamics of the string operator $X_j$ is also thoroughly studied. In Sec.~\ref{SecIV}, we study the reduced dynamics of two nearest-neighbor spins in detail. Conclusions are drawn in Sec.~\ref{SecV}.
\section{Model, initial state, and time-evolved state}\label{SecII}
\subsection{Model and diagonalization}
\par The ferromagnetic quantum Ising ring with $N$ sites is described by the Hamiltonian (set $\hbar=1$)
\begin{eqnarray}\label{HQIM}
H_{\mathrm{QIM}}=-\sum^N_{j=1}(\sigma^x_j\sigma^x_{j+1}+g\sigma^z_j),
\end{eqnarray}
where $\vec{\sigma}_j$ is the Pauli operator for spin-$j$ and $g\geq0$ is a transverse field along the $z$ axis. We assume that $N$ is even and use the periodic boundary conditions with $\vec{\sigma}_1=\vec{\sigma}_{N+1}$, which ensures the translational invariance of the spin chain. Besides the circular symmetry, the homogeneous ring also holds inversion symmetry about any site. In the thermodynamic limit, the model exhibits a quantum phase transition at the critical point $g_{\mathrm{c}}=1$ between the ordered phase ($0\leq g<1$) and the disordered phase ($g>1$). To introduce the notations used below, we first briefly review the diagonalization of $H_{\mathrm{QIM}}$.
\par $H_{\mathrm{QIM}}$ can be mapped onto a spinless fermion model through the standard Jordan-Wigner transformation~\cite{Lieb1961}
\begin{eqnarray}\label{JWT}
\sigma^-_j\equiv\frac{\sigma^x_j-i\sigma^y_j}{2}=T_j c_j,~\sigma^+_j=T_j c^\dag_j,~\sigma^z_j=2c^\dag_j c_j-1,
\end{eqnarray}
where the $c^\dag_j$'s are fermionic creation operators and $T_j=\prod^{j-1}_{l=1}(-\sigma^z_l)=e^{i\pi\sum^{j-1}_{l=1}c^\dag_l c_l}$ is the JW string operator. Due to the presence of the boundary term $-\sigma^x_N\sigma^x_1$, applying Eq.~(\ref{JWT}) in Eq.~(\ref{HQIM}) does not lead to simple cyclic boundary conditions for the fermionic chain:
\begin{eqnarray}\label{HF}
-\sigma^x_{N}\sigma^x_{1}=(c^\dag_Nc_1+c^\dag_1c_N+c^\dag_Nc^\dag_1+c_1c_N)T_{N+1},\nonumber
\end{eqnarray}
where $T_{N+1}=e^{i\pi\sum^{N}_{l=1}c^\dag_l c_l}$ is the fermion parity operator. It is easy to check that $T_{N+1}$ is conserved and has eigenvalues $\pm 1$. Accordingly, $H_{\mathrm{QIM}}$ can be separately diagonalized in two subspaces with even ($T_{N+1}=1$) and odd ($T_{N+1}=-1$) fermion parities. We now define two projection operators
\begin{eqnarray}\label{PP}
 P_+=\frac{1}{2}(1+ T_{N+1}),~P_-=\frac{1}{2}(1- T_{N+1}),
\end{eqnarray}
which project respectively onto the even and odd subspace and satisfy
\begin{eqnarray}\label{PPP}
P_++P_-=1,~P_+P_-=P_-P_+=0.
\end{eqnarray}
An operator is said to be even (odd) if it can be expressed as a combination of products of even (odd) numbers of JW fermion operators. For example, the transverse spin $\sigma^z_j=2c^\dag_j c_j-1$ is even; while the longitudinal spin $\sigma^x_j=\prod^{j-1}_{l=1}(1-2c^\dag_l c_l)(c_j+c^\dag_j)$ is odd. An odd operator $A_{\mathrm{o}}$ changes the fermion parity and can be written as
\begin{eqnarray}\label{Ao}
A_{\mathrm{o}}=(P_++P_-)A_{\mathrm{o}}(P_++P_-)=\sum_{\sigma=\pm}P_\sigma A_{\mathrm{o}}P_{-\sigma}.
\end{eqnarray}
Similarly, an even operator $A_{\mathrm{e}}$ preserves the fermion parity and has the form
\begin{eqnarray}\label{Ae}
A_{\mathrm{e}}=(P_++P_-)A_{\mathrm{e}}(P_++P_-)=\sum_{\sigma=\pm}P_\sigma A_{\mathrm{e}}P_{\sigma}.
\end{eqnarray}
The fermionic Hamiltonian after the JW transformation,  $H_{\mathrm{F}}$, is obviously even. With the help of Eq.~(\ref{PP}), $H_{\mathrm{F}}$ can be expressed as
\begin{eqnarray}\label{HFPP}
H_{\mathrm{F}} &=& P_+H_{\mathrm{F},+}P_++P_-H_{\mathrm{F},-}P_-,\nonumber\\
H_{\mathrm{F},\sigma}&=&- \sum^{N-1}_{j=1} (c^\dag_j c_{j+1}+c^\dag_j c^\dag_{j+1}+\mathrm{H.c.})-2g\sum^N_{j=1}  c^\dag_jc_j + gN\nonumber\\
&& +\sigma  (c^\dag_Nc_1+c^\dag_Nc^\dag_1+\mathrm{H.c.}).
\end{eqnarray}
We now define $c_{N+1}\equiv -\sigma c_1$, then  $H_{\mathrm{F},\sigma}$ can be diagonalized via the following Fourier transformations
\begin{eqnarray}\label{cj}
c_{j}&=& \frac{e^{i\pi/4}}{\sqrt{N}}\sum_{k\in K_\sigma}e^{ikj}c_{k\sigma},~~c_{k\sigma}= \frac{e^{-i\pi/4}}{\sqrt{N}}\sum^N_{j=1}e^{-ikj}c_{j}
\label{cjsigma}
\end{eqnarray}
as
\begin{eqnarray}\label{HFpm}
H_{\mathrm{F},+}&=&\sum_{ k\in K'_+}H_{k+},\nonumber\\
H_{\mathrm{F},-}&=&\sum_{ k\in K'_-}H_{k-}+\frac{1}{2}(H_{-\pi}+H_{0}),
\end{eqnarray}
where $K_\sigma$ is the set of the allowed wave numbers in the $\sigma$-sector:
\begin{eqnarray}
K_+&=&\left\{-\pi+\frac{\pi}{N},-\pi+\frac{3\pi}{N}\cdots,-\frac{\pi}{N},\frac{\pi}{N},\cdots,\pi-\frac{\pi}{N}\right\},\nonumber\\
\label{K+}
K_-&=&\left\{-\pi,-\pi+\frac{2\pi}{N}\cdots,0,\frac{2\pi}{N},\cdots,\pi-\frac{2\pi}{N}\right\},
\label{K-}
\end{eqnarray}
and $K'_\sigma$ is the subset of $K_\sigma$ obtained by keeping only the positive elements. The mode Hamiltonians in Eq.~(\ref{HFpm}) read
\begin{eqnarray}
H_{k,\sigma}&=& -2(c^\dag_{k\sigma},c_{-k,\sigma})\left(
                                                     \begin{array}{cc}
                                                       \cos k+g &\sin k \\
                                                       \sin k & -\cos k-g \\
                                                     \end{array}
                                                   \right) \left(
                                                       \begin{array}{c}
                                                         c_{k\sigma} \\
                                                         c^\dag_{-k,\sigma} \\
                                                       \end{array}
                                                     \right),\nonumber\\
H_{-\pi}&=&2(1-g)(2c^\dag_{-\pi }c_{-\pi }-1),\nonumber\\
H_{0}&=& -2(1+g)(2c^\dag_{0 }c_{0 }-1).
\end{eqnarray}
\par In the special case of $g=0$, the quantum Ising ring reduces to the classical Ising ring whose ground states are simply the two degenerate ferromagnetic states,
\begin{eqnarray}\label{RL}
|R\rangle&=&|\rightarrow\cdots\rightarrow\rangle=\left(\frac{1}{\sqrt{2}}\right)^N\prod^N_{j=1}(1+ c^\dag_j)|\mathrm{vac}\rangle,\nonumber\\
|L\rangle&=&|\leftarrow\cdots\leftarrow\rangle=\left(\frac{1}{\sqrt{2}}\right)^N\prod^N_{j=1}(1- c^\dag_j)|\mathrm{vac}\rangle,
\end{eqnarray}
where $|\mathrm{vac}\rangle$ is the common vacuum of $c_j,~\forall j$. Although having a product form in the spin representation, the two states $|R\rangle$ and $|L\rangle$ do not admit simple forms in the real-space fermion representation. More importantly, neither $|R\rangle$ nor $|L\rangle$ has a definite fermion parity, as can be seen from the right-hand side of Eq.~(\ref{RL}).
\par Fortunately, it is shown in Ref.~\cite{PRE2020} that $|R\rangle$ and $|L\rangle$ can be written as equally weighted linear superpositions of the two ground states in the momentum space with $g=0$:
\begin{eqnarray}\label{RLG}
|R\rangle&=&\frac{1}{\sqrt{2}}(|G_+\rangle+e^{-i\frac{\pi}{4}}|G_-\rangle),\nonumber\\
|L\rangle&=&\frac{1}{\sqrt{2}}(|G_+\rangle-e^{-i\frac{\pi}{4}}|G_-\rangle).
\end{eqnarray}
Here,
\begin{eqnarray}\label{Gpm}
|G_- \rangle&=&|0\rangle\prod_{k\in K'_-}\left(\sin\frac{k}{2}|\mathrm{vac}\rangle_{k-}+\cos\frac{k}{2}|k,-k\rangle_{-}\right),\nonumber\\
|G_+ \rangle&=&\prod_{ k\in K'_+}\left(\sin\frac{k}{2}|\mathrm{vac}\rangle_{k+}+\cos\frac{k}{2}|k,-k\rangle_{+}\right),
\end{eqnarray}
are the two degenerate ground states of the classical Ising ring obtained from Eq.~(\ref{HFpm}), with $|\mathrm{vac}\rangle_{k\sigma}$ the vacuum state for both $c_{k\sigma}$ and $c_{-k\sigma}$ and $|k,-k\rangle_\sigma\equiv c^\dag_{k\sigma}c^\dag_{-k\sigma}|\mathrm{vac}\rangle_{k\sigma}$ the doubly occupied state for mode $k>0$. The state $|0\rangle\equiv c^\dag_{k=0}|\mathrm{vac}\rangle_{k=0,-}$ is the singly occupied state with zero momentum in the odd sector. Note that $|G_+ \rangle$ ($|G_- \rangle$) has an even (odd) fermion parity.
\subsection{Initial state and time-evolved state}
\par Although the ground state of $H_{\mathrm{QIM}}$ is twofold degenerate for $g=0$, to probe interesting non-equilibrium dynamics of the system we choose as the initial state the ferromagnetic state $|R\rangle$ that breaks the $Z_2$ symmetry,
\begin{eqnarray}
|\psi_0\rangle=|R\rangle.
\end{eqnarray}
After a sudden quench of the transverse field from $g_{\mathrm{i}}=0$ to $g_{\mathrm{f}}=g$, the time evolution of the system is governed by $H_{\mathrm{QIM}}$ with finite $g$.
Note that $|\psi_0\rangle$ is invariant under both lattice translation and inversion, which guarantees that the time evolved state $e^{-iH_{\rm{QIM}}t}|\psi_0\rangle$ preserves the same symmetry. As a result, expectation values of the spin operators in the time-evolved state must satisfy the following properties:
\begin{eqnarray}\label{sym}
\langle\sigma^\alpha_i\sigma^\beta_{j}\cdots\rangle_t&=&\langle\sigma^\alpha_{i+l}\sigma^\beta_{j+l}\cdots\rangle_t,\nonumber\\
\langle\sigma^\alpha_i\sigma^\beta_{i+l}\sigma^\gamma_{i+m}\cdots\rangle_t&=&\langle\cdots\sigma^\gamma_{j-m}\sigma^\beta_{j-l}\sigma^\alpha_j\rangle_t.
\end{eqnarray}
\par Using Eqs.~(\ref{RLG}) and (\ref{Gpm}), the time-evolved state is simply obtained by evolving each mode state, giving
\begin{eqnarray}
|\psi(t)\rangle&=&e^{-iH_{\rm{QIM}}t}|\psi_0\rangle\nonumber\\
&=&\frac{1}{\sqrt{2}}(|\phi_+(t)\rangle+e^{-i\frac{\pi}{4}}|\phi_-(t)\rangle),
\end{eqnarray}
where
\begin{eqnarray}
|\phi_+(t)\rangle&=&e^{-iH_{\mathrm{F},+}t}|G_+\rangle=\prod_{k\in K'_+}|\chi_k\rangle_{+},\nonumber\\
|\phi_-(t)\rangle&=&e^{-iH_{\mathrm{F},-}t}|G_-\rangle=e^{i2t}|0\rangle\prod_{k\in K'_-}|\chi_k\rangle_{-},
\end{eqnarray}
with
\begin{eqnarray}
|\chi_k\rangle_{\sigma}&\equiv& \left[u^{(\sigma)}_k(t)+v^{(\sigma)}_k(t)c^\dag_{k,\sigma}c^\dag_{-k,\sigma}\right]|\mathrm{vac}\rangle_{k\sigma}.
\end{eqnarray}
Since the Hamiltonian is time-independent, the coefficients $u^{(\sigma)}_k$'s and $v^{(\sigma)}_k$'s can be calculated analytically as
\begin{eqnarray}
u^{(\sigma)}_k(t)&=&\sin\frac{k}{2}\left[\cos\Lambda_kt+2i(1-g)\frac{\sin\Lambda_k t}{\Lambda_k}\right],\nonumber\\
v^{(\sigma)}_k(t)&=&\cos\frac{k}{2}\left[\cos\Lambda_kt+2i(1+g)\frac{\sin\Lambda_k t}{\Lambda_k}\right],
\end{eqnarray}
where
\begin{eqnarray}\label{Lambdak}
\Lambda_k&=&2\sqrt{g^2+2g\cos k+1},
\end{eqnarray}
is the single-particle dispersion.
\par The expectation value of an even operator $A_{\mathrm{e}}$ in the time-evolved state $|\psi(t)\rangle$ can be separately calculated in each parity sector:
\begin{eqnarray}\label{Aet}
\langle A_{\mathrm{e}} \rangle_t&\equiv& \langle\psi(t)|A_{\mathrm{e}}|\psi(t)\rangle\nonumber\\
&=&\frac{1}{2}\sum_{\sigma=\pm} \langle\phi_{\sigma}(t)|P_{\sigma}A_{\mathrm{e}}P_\sigma|\phi_\sigma(t)\rangle.
\end{eqnarray}
In most cases of interest, the expectation values on the right-hand side of Eq.~(\ref{Aet}) can be calculated through standard free-fermion techniques~\cite{Lieb1961}.
\par Since $|\phi_+(t)\rangle$ and $|\phi_-(t)\rangle$ have distinct fermion parities, the expectation value of a generic odd operator $A_{\mathrm{o}}$ in the time-evolved state $|\psi(t)\rangle$ does not vanish:
\begin{eqnarray}\label{Aot}
\langle A_{\mathrm{o}} \rangle_t&\equiv& \langle\psi(t)|A_{\mathrm{o}}|\psi(t)\rangle\nonumber\\
&=&\frac{1}{2}\sum_{\sigma=\pm}e^{i\sigma\frac{\pi}{4}}\langle\phi_{-\sigma}(t)|P_{-\sigma}A_{\mathrm{o}}P_\sigma|\phi_\sigma(t)\rangle.
\end{eqnarray}
The evaluation of the matrix element on the right-hand side of Eq.~(\ref{Aot}) is usually a difficult task.
\section{Reduced dynamics of a single spin}\label{SecIII}
\subsection{Dynamics of the transverse and longitudinal magnetizations}
Because of the translational invariance of the time-evolved state, we can consider the reduced density matrix of an arbitrary site, say site $1$,
\begin{eqnarray}
\rho^{(1)}(t)=\frac{1}{2}(1+\langle\sigma^x_1\rangle_t\sigma^x_1+\langle\sigma^y_1\rangle_t\sigma^y_1+\langle\sigma^z_1\rangle_t\sigma^z_1).
\end{eqnarray}
We thus need to calculate the dynamics of both the transverse ($\langle\sigma^z_1\rangle_t$) and longitudinal magnetizations ($\langle\sigma^{x,y}_1\rangle_t$).
\par  Since $\sigma^z_1$ is an even operator, the dynamics of the transverse magnetization can be calculated using Eq.~(\ref{Aet}). From the relation $\sigma^z_1=2c^\dag_1c_1-1$ and writing $c_1$ and $c^\dag_1$ in the momentum space, it can be easily shown that
\begin{eqnarray}\label{Sztfinal}
\langle\sigma^z_1\rangle_t&=&\frac{1}{N}\left( 2\sum_{\sigma=\pm}\sum_{k \in K'_\sigma} |v^{(\sigma)}_k(t)|^2+1\right)-1.
\end{eqnarray}
\par The calculation of the longitudinal magnetization dynamics is less straightforward since $\sigma^{x}_1$ and $\sigma^{y}_1$ are odd operators and change the fermion parity. As a result, we have to evaluate the inner product between two states within distinct parity sectors. This difficulty has been noticed by several authors~\cite{Damski2020,Rossini2020,PRA1971,JSM2012} and the longitudinal magnetization dynamics is usually treated by exact diagonalization for small rings~\cite{Yu2019,Damski2020,Rossini2020} or in the thermodynamic limit~\cite{JSM2012,Eisler2018}.
\par In general, the dynamics of the local fermion operator $c_j$, or equivalently the string operator $\prod^{j-1}_{l=1} (-\sigma^z_l) \sigma^-_j$, can be calculated from Eq.~(\ref{Aot}) as
\begin{eqnarray}\label{lengthy}
&&\langle c_j\rangle_t=(-1)^{j-1}\langle \prod^{j-1}_{l=1} \sigma^z_l \sigma^-_j\rangle_t\nonumber\\
&=&\frac{1}{2\sqrt{N}} \langle\phi_+(t)|  c_{0-}|\phi_-(t)\rangle\nonumber\\
&&+\frac{1}{2\sqrt{N}}\langle\phi_+(t)|\sum_{k\in K'_-}(e^{ikj} c_{k-}+e^{-ikj} c_{-k,-})|\phi_-(t)\rangle\nonumber\\
&&+\frac{i}{2\sqrt{N}}\langle\phi_-(t)|\sum_{p\in K'_+}(e^{ipj} c_{p+}+e^{-ipj} c_{-p,+})|\phi_+(t)\rangle.\nonumber\\
\end{eqnarray}
It is shown in Ref.~\cite{PRE2020} that the typical inner product appearing on the right-hand side of the Eq.~(\ref{lengthy}) can be expressed as the Pfaffian of an appropriate matrix. Thus, $\langle c_j\rangle_t$ can be evaluated in an semi-analytical way through efficient numerical computation of the Pfaffian~\cite{PfWim}. Once $\langle c_j\rangle_t$ is obtained, the evolved longitudinal magnetizations are simply
\begin{eqnarray}
 \langle\sigma^x_1\rangle_t&=&\langle c_1\rangle_t+\langle c_1\rangle_t^*,\nonumber\\
  \langle\sigma^y_1\rangle_t&=&i(\langle c_1\rangle_t-\langle c_1\rangle_t^*).
\end{eqnarray}
\par We are also interested in the dynamics of the string operator $X_j\equiv \prod^{j-1}_{l=1}\sigma^z_l\sigma^x_j$:
\begin{eqnarray}
\langle X_j\rangle_t=(-1)^{j-1}[\langle c_j\rangle_t+\langle c_j\rangle^*_t].
\end{eqnarray}
Note that $X_j$ is a local operator of range $j$, and hence can generally relax locally after a quantum quench~\cite{JSM2016}.
\begin{figure}
\includegraphics[width=.50\textwidth]{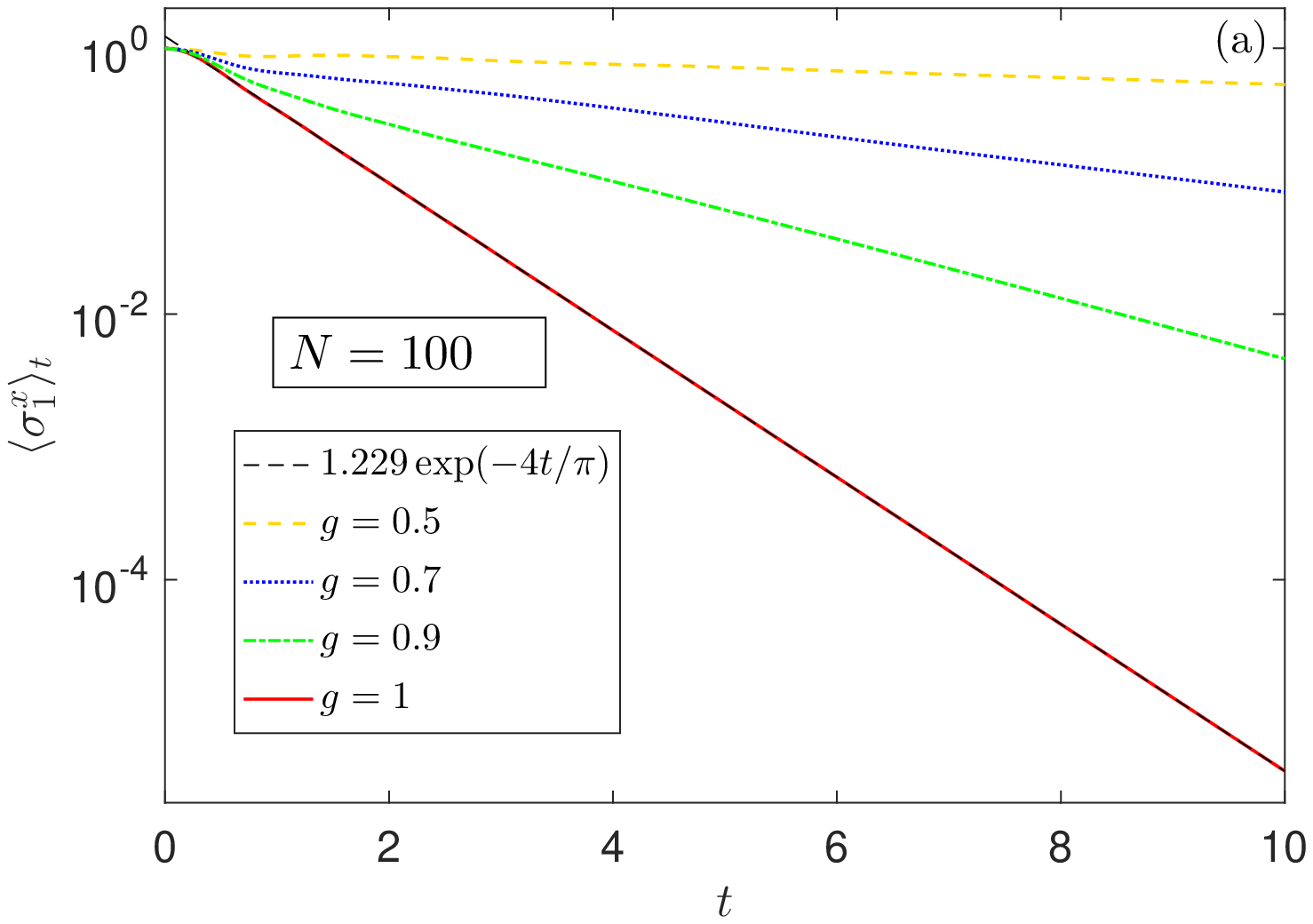}
\includegraphics[width=.50\textwidth]{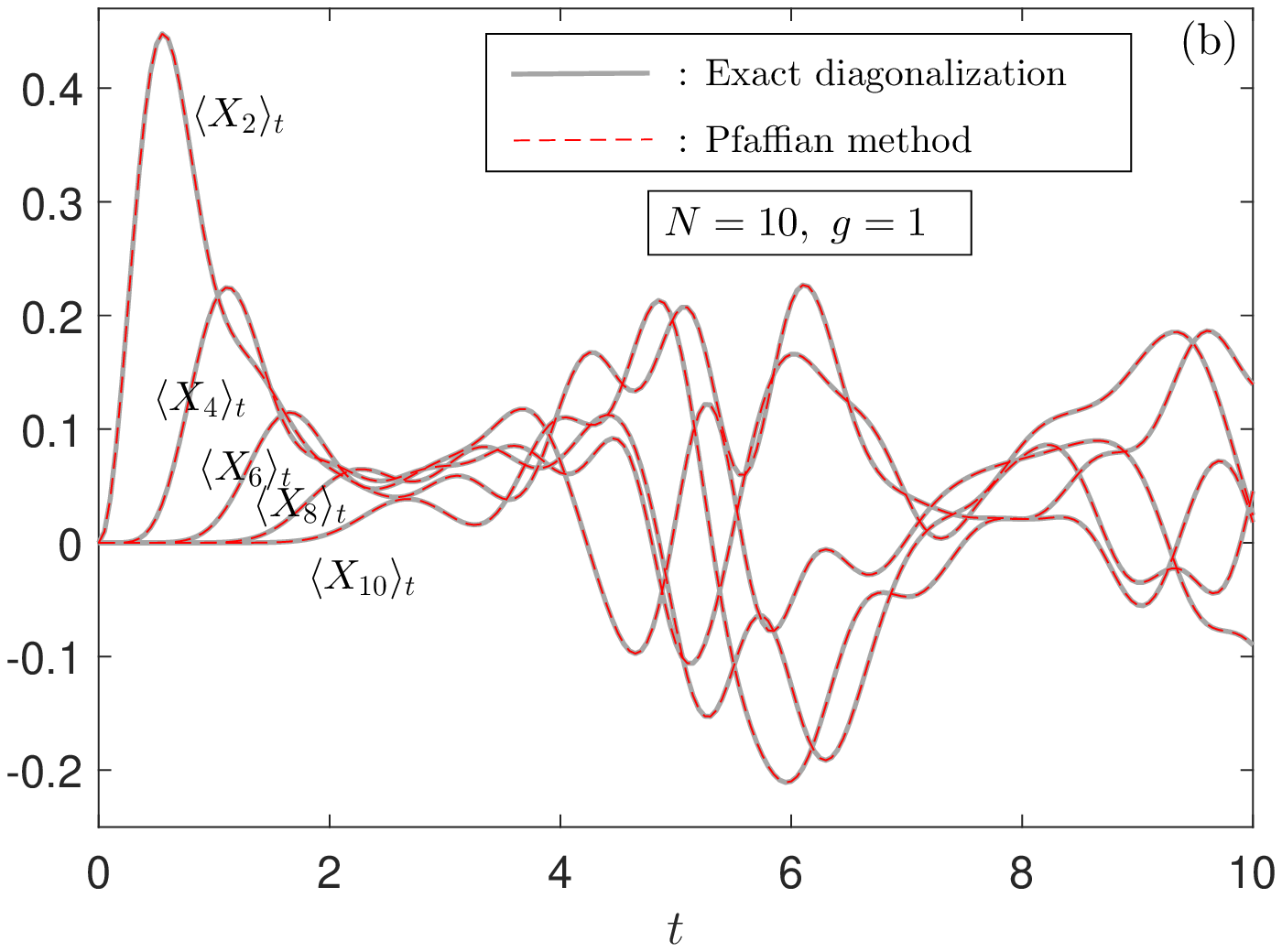}
\caption{(a) Dynamics of the longitudinal magnetization $\langle\sigma^x_1\rangle_t$ after quenches within the ordered phase. The numerical simulations are performed for a large quantum Ising ring with $N=100$ sites. It can be seen that $\langle\sigma^x_1\rangle_t$ decays exponentially at long times, confirming the asymptotic behavior derived in Ref.~\cite{JSM2012} for $g<1$ and $N\to\infty$. The dashed black curve shows a fit for $g=1$ discovered in Ref.~\cite{Rossini2020} through exact diagonalization. (b) Dynamics of the string operator $\langle X_j\rangle_t$ after a quench to $g=1$ for a small ring with $N=10$ sites. The gray and dashed red curves show the results obtained by exact diagonalization and the Pfaffian method, respectively.}
\label{Fig1}
\end{figure}
\subsection{Numerical results}
\par To see the power of our method, we plot in Fig.~\ref{Fig1}(a) the longitudinal magnetization $\langle\sigma^x_1\rangle_t$ after quenches within the ordered phase ($g\leq 1$) for a large quantum Ising ring with $N=100$ spins. The system size we choose is large enough to faithfully capture the dynamical behavior in the thermodynamic limit and the numerical simulations can be performed on a personal computer. Our numerical results confirm the asymptotic exponential decay of $\langle\sigma^x_1\rangle_t$ at long times that was derived in Ref.~\cite{JSM2012} for $g<1$ and $N\to\infty$, as well as a recently discovered correction to the prefactor for $g=1$~\cite{Rossini2020} [dashed black curve in Fig.~\ref{Fig1}(a)]. To further verify the validity of our method, we calculate the dynamics of the string operator $X_j$ in a small ring with $N=10$ sites by using both exact diagonalization and the Pfaffian method. As can be seen from Fig.~\ref{Fig1}.(b), the differences between the two results are hardly visible.
\subsubsection{Dynamics of the string operator $X_j$}
\begin{figure}
\includegraphics[width=.53\textwidth]{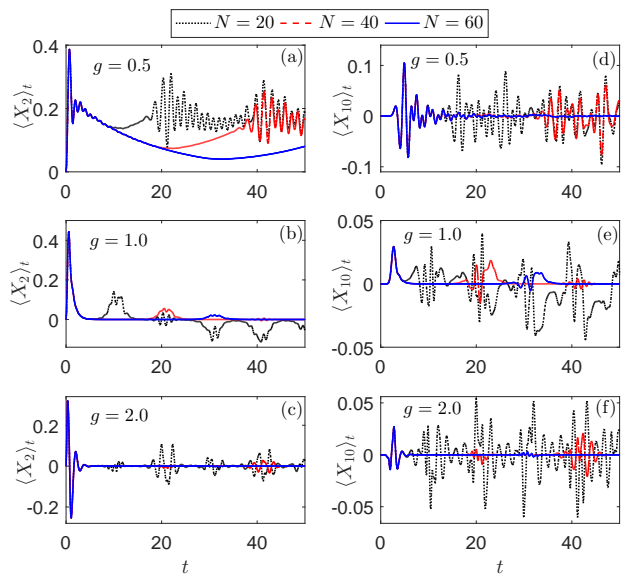}
\caption{Dynamics of $\langle X_2\rangle_t$ [(a)-(c)] and $\langle X_{10}\rangle_t$ [(d)-(f)] after quenches to various values of $g$ and for different system sizes.}
\label{Fig2}
\end{figure}
\par It is apparent that $\langle X_j\rangle_{t=0}$ vanishes for $j>1$ since the transverse spin $\sigma^z_l$ flips the state $|\rightarrow\rangle_l$ to $|\leftarrow\rangle_l$. Qualitatively, we expect that a longer period of time $t$ is needed for $\langle X_j\rangle_{t}$ with a larger $j$ to acquire a value significantly different zero. This is already evident from Fig.~\ref{Fig1}.(b), where we see that $\langle X_j\rangle_t$ reaches its first maximum at a time that increases monotonically with increasing $j$.
\par To understand the behavior of $\langle X_j\rangle_{t}$ for different system sizes and string lengths, we plot in Fig.~\ref{Fig2} $\langle X_2\rangle_{t}$ (left panels) and $\langle X_{10}\rangle_{t}$ (right panels) for various values of $g$ and $N$. For all quenches considered, $\langle X_2\rangle_{t}$ rapidly increases to its first maximum, followed by a sudden drop over short times scales. The behavior of $\langle X_2\rangle_{t}$ after reaching its first maximum is similar to that of $\langle\sigma^x_1\rangle_t$~\cite{Rossini2020,PRE2020}. For quenches within the ordered phase $\langle X_2\rangle_{t}$ decays exponentially for large enough $N$. For quenches to the disordered phase $\langle X_2\rangle_{t}$ decays more abruptly from its maximum to negative values. The right three panels of Fig.~\ref{Fig2} show the results for $\langle X_{10}\rangle_t$. As expected, $\langle X_{10}\rangle_t$ reaches its first maxima over a longer time scale compared to $\langle X_{2}\rangle_t$. For $g=0.5$ and large enough $N$, $\langle X_{10}\rangle_t$ experiences a period of oscillation after the first maximum is reached and then approaches a nearly vanishing value with minor oscillations [Fig.~\ref{Fig2}(d)]. For $g=1$, $\langle X_{10}\rangle_t$ decays to zero from the first maximum in the thermodynamic limit.
\par Quantum quench to the critical point is an important case and deserves further investigation. It is intriguing to observe that there is a (partial) revival of $\langle X_{j}\rangle_t$ around $t=N/2$ [Figs.~\ref{Fig2}(b) and (e)]. Such revivals are actually typical and ubiquitous in the dynamics of local observables after a quench to the critical point~\cite{Rossini2020,PRE2020,Cardy2014,Najafi2017,JSM2018}. As shown by Cardy~\cite{Cardy2014}, the revivals generally occur at $t=Nm/2,~m\in Z$ for a critical circular spin chain whose conformal field theory having central charge $c<1$. Our numerical results are thus consistent with this picture since the Ising conformal field theory has central charge $c=1/2$. It is also demonstrated that correlation functions of local observables within a subsystem of length $j$ become stationary for times $t$ such that $j/2<t<(N-j)/2$~\cite{Cardy2014}. By investigating the behavior of $\langle X_{10}\rangle_t$ (which involves correlations within a string of length $j=10$) in Fig.~\ref{Fig2}(e) we see that it nearly collapse within the interval $t\in[5,25]$ for $N=60$, again confirming the above picture.
\begin{figure}
\includegraphics[width=.53\textwidth]{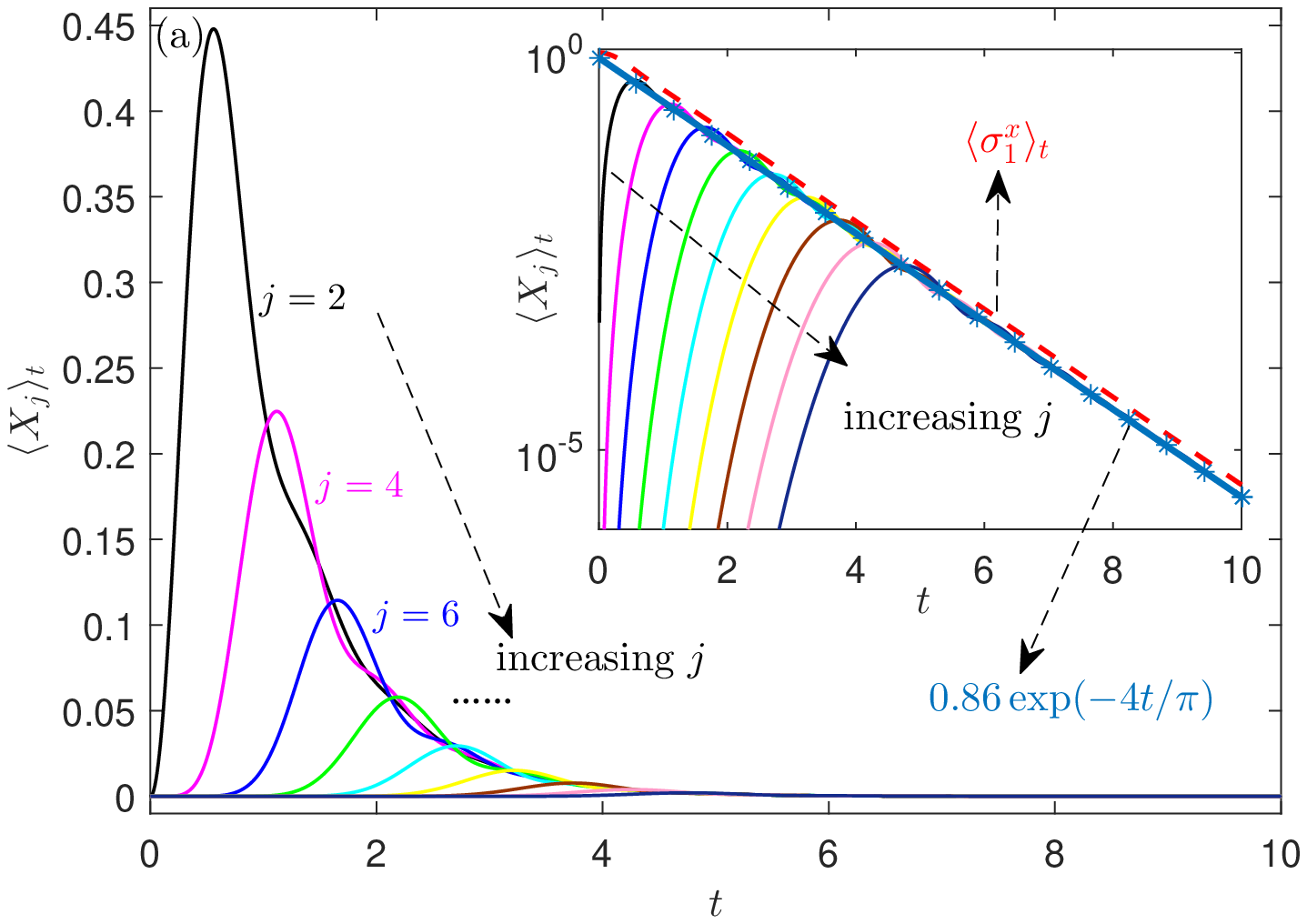}
\includegraphics[width=.53\textwidth]{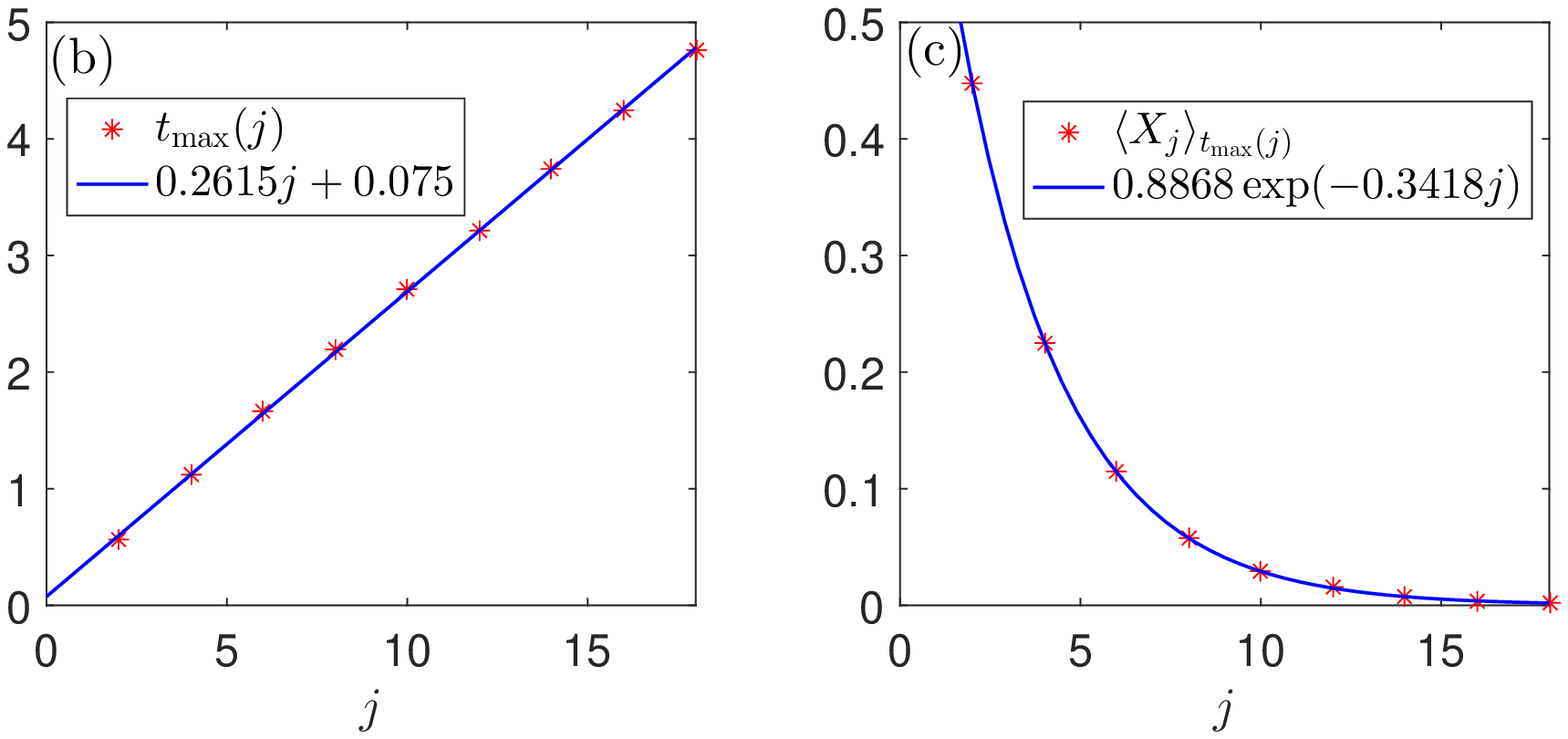}
\caption{(a) Dynamics of $\langle X_j\rangle_t$ for $j=2,4,\cdots,18$ after a quench to the critical point $g=1$ in a ring with $N=60$ sites. The inset shows a semi-log plot of the same result. At long times the $\langle X_j\rangle_t$'s collapse into a $j$-independent exponential decay of the form $0.86 e^{-4t/\pi}$. The longitudinal magnetization dynamics (dashed red line) is also presented for comparison. (b) The time at which $\langle X_j\rangle_t$ reaches its first maximum, $t_{\max}(j)$, as a function of $j$ (red stars). The blue curve is a linear regression. (c) The first maximum of $\langle X_j\rangle_t$ as a function of $j$ (red stars) and an exponential fitting (blue curve). }
\label{Fig3}
\end{figure}
\par It is interesting to note that in all cases the short-time dynamics of $\langle X_{j}\rangle_t$ is insensitive to the system size $N$. For fixed $j$, simulations on a system with $N\approx 2j$ sites is already sufficient to accurately capture the very time at which the first maximum of $\langle X_{j}\rangle_t$ appear in the thermodynamic limit. Figure~\ref{Fig3}(a) shows the evolution of $\langle X_j\rangle_t$ after a quench to the critical point $g=1$ for a ring with $N=60$ sites. Results for $j=2,4,\cdots,18$ are shown since the $\langle X_{j}\rangle_t$'s are vanishingly small for $j>18$. The system size is large enough to observe a universal $j$-independent exponential decay of $\langle X_{j}\rangle_t$ at long times over the considered time scale [inset of Fig.~\ref{Fig3}(a)]. An exponential fitting of the data in the long time regime ($t\geq 4$) gives an asymptotic form $\sim 0.86 e^{-4t/\pi}$. Although the decay rate $4/\pi$ is the same as that of $\langle\sigma^x_1\rangle$, we observe a smaller prefactor for $\langle X_j\rangle_t$ with $j>1$ [see Fig.~\ref{Fig1}(a) and the dashed red curve in the inset of Fig.~\ref{Fig3}(a)].
\par Figure~\ref{Fig3}(b) shows the time at which $\langle X_j\rangle_t$ reaches its first maximum, $t_{\max}(j)$, as a function of string length, $j$. It can be seen that $t_{\max}(j)$ scales linearly with $j$. On the other hand, the first maxima $\langle X_{j}\rangle_{t_{\max}(j)}$ is found to decay exponentially with increasing $j$ [Fig.~\ref{Fig3}(c)].
\subsubsection{Magnetization and purity dynamics}
\begin{figure}
\includegraphics[width=.50\textwidth]{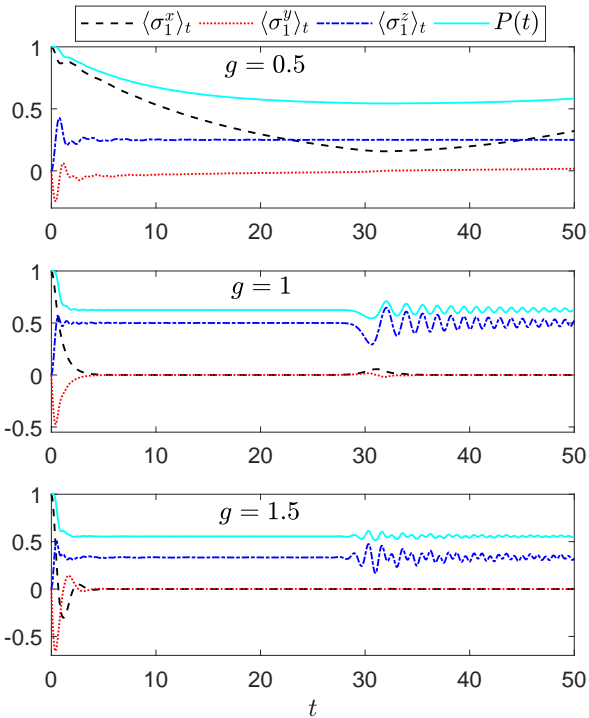}
\caption{Dynamics of the magnetizations $\langle\sigma^x_1\rangle_t$, $\langle\sigma^y_1\rangle_t$, $\langle\sigma^z_1\rangle_t $, and the corresponding purity $P(t)$ after several quenches in a ring with $N=60$ sites.}
\label{Fig4}
\end{figure}
\par Let us now discuss the magnetization dynamics. Given the reduced density matrix $\rho^{(1)}(t)$, we also monitor the purity dynamics of an arbitrary spin,
\begin{eqnarray}
P(t)=\frac{1}{2}(1+\langle\sigma^x_1\rangle^2_t+\langle\sigma^y_1\rangle^2_t+\langle\sigma^z_1\rangle^2_t).
\end{eqnarray}
Figure~\ref{Fig4} shows the time evolutions of $\langle\sigma^x_1\rangle_t$, $\langle\sigma^y_1\rangle_t$, $\langle\sigma^z_1\rangle_t $ and the purity $P(t)$ for quenches to $g=0.5$, $1$, and $1.5$. For the quench within the ordered phase ($g=0.5$), the purity $P(t)$ shows a similar trend to the longitudinal magnetization $\langle\sigma^x_1\rangle_t$ since both $\langle\sigma^y_1\rangle_t$ and $\langle\sigma^z_1\rangle_t$ reach their steady values after $t\approx5$. For quenches to $g\geq 1$, both $\langle\sigma^x_1\rangle_t$ and $\langle\sigma^y_1\rangle_t$ approach zero rapidly and $P(t)$ is determined solely by the transverse magnetization $\langle\sigma^z_1\rangle_t $, which shows an oscillating behavior after $t\approx 30=N/2$ (the `revival' due to finite-size effect, see also Ref.~\cite{Rossini2020}).
\begin{figure}
\includegraphics[width=.50\textwidth]{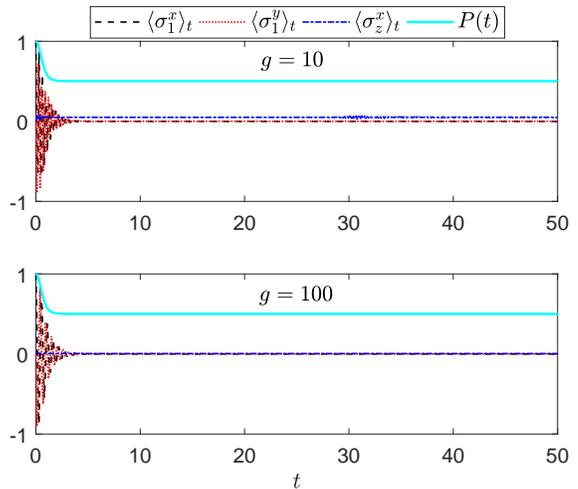}
\caption{Dynamics of the magnetizations $\langle\sigma^x_1\rangle_t$, $\langle\sigma^y_1\rangle_t$, and $\langle\sigma^z_1\rangle_t $, and the corresponding purity $P(t)$ after quenches to the strong field regime with $g=10$ (upper panel) and $g=100$ (lower panel). Numerical simulations are performed for $N=60$.}
\label{Fig5}
\end{figure}
\par It is interesting to see what happens when the field is quenched to the strong field regime with $g\gg1$. From Fig.~\ref{Fig5} we see that for quenches to the extremely disordered phase all the three components of the single-spin polarization tend to vanish after intensive initial oscillations. As a result, the purity $P(t)$ keeps a steady value of $\sim0.5$, indicating that the single-spin reaches a maximally mixed state under the influence of the strong transverse field. It is qualitatively reasonable that as time evolves strong transverse fields will destroy the longitudinal magnetizations. To understand the behavior of $\langle\sigma^z_1\rangle_t$ for large $g$, let us consider the thermodynamics limit $N\to\infty$, in which Eq.~(\ref{Sztfinal}) becomes (using $\frac{2}{N}\sum_{k\in K'_\sigma}\to\frac{1}{\pi}\int^\pi_0dk$)
\begin{eqnarray}\label{SzTL}
 \lim_{N\to\infty}\langle\sigma^z_1\rangle_t=\frac{8g}{\pi}\int^\pi_0dk  \sin^2k\frac{1-\cos 2\Lambda_kt}{\Lambda^2_k}.
\end{eqnarray}
For $g\gg1$ the denominator can be approximated as $\Lambda^2_k\approx 4g^2$, giving
\begin{eqnarray}\label{SzTL}
 \lim_{N\to\infty}\langle\sigma^z_1\rangle_t\approx \frac{2}{g\pi}\int^\pi_0dk  \sin^2k(1-\cos 2\Lambda_kt),
\end{eqnarray}
which tends to zero as $1/g$ when $g\to\infty$.
\par Another interesting regime for the quench is in the vicinity of the critical point. It was previously derived in Ref.~\cite{JSM2012} that in the thermodynamic limit $N\to\infty$ the longitudinal magnetization decays exponentially after a quench to $g<1$ according to
\begin{eqnarray}
\lim_{N\to\infty}\langle\sigma^x_1\rangle_{t}=A(g) e^{-\gamma(g)t},~g<1
\end{eqnarray}
where
\begin{eqnarray}
\label{Ag}
A(g)&=&\frac{1}{\sqrt{2}}\sqrt{1+\sqrt{1-g^2}},\\
\label{rg}
\gamma(g)&=&\frac{4g}{\pi}\int^\pi_0\frac{\sin k}{\Lambda_k}\ln\frac{2(1+g\cos k)}{\Lambda_k}.
\end{eqnarray}
It is easy to check that
\begin{eqnarray}
\lim_{g\to 1^-}\gamma(g)=\frac{4}{\pi}.
\end{eqnarray}
In Ref.~\cite{Rossini2020} it was further shown through exact diagonalization in a ring with $N=24$ sites that around $g=1$ the decay function $\gamma(t)$ can be approximated as
\begin{eqnarray}\label{Gg}
\gamma(g)=\frac{4}{\pi}-2\sqrt{2(1-g)}+O(1-g),~g<1.
\end{eqnarray}
\begin{figure}
\includegraphics[width=.54\textwidth]{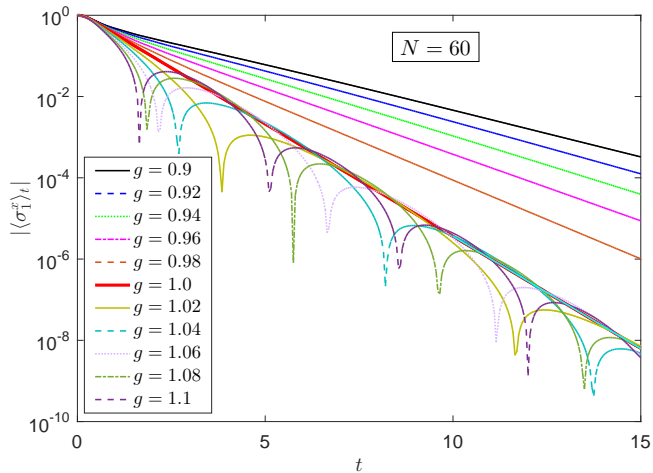}
\caption{Dynamics of the absolute value of $\langle\sigma^x_1\rangle_t$ after quenches into the vicinity of the critical point $g=1$ for a ring with $N=60$ sites.}
\label{Fig6}
\end{figure}
In addition, it is confirmed numerically that $A(g\to1^{-})=1/\sqrt{2}$ does not provide the correct prefactor for a quench to exactly $g=1$~\cite{Rossini2020} [see also Fig.~\ref{Fig1}(a)]. It is thus desirable to test these behaviors in finite-size but large systems using our method.
\par Figure~\ref{Fig6} shows time evolution of the absolute value of $\langle\sigma^x_1\rangle_t$ after quenches to $g$'s lying within the interval $[0.9,1.1]$. As expected, $\langle\sigma^x_1\rangle_t$ is always positive and decays exponentially for $g\leq 1$ over the considered time scale. Although $\langle\sigma^x_1\rangle_t$ acquires negative values for $g>1$, the envelope of $|\langle\sigma^x_1\rangle_t|$ still evolve roughly along the curve corresponding to $g=1$.
\par
\begin{figure}
\includegraphics[width=.53\textwidth]{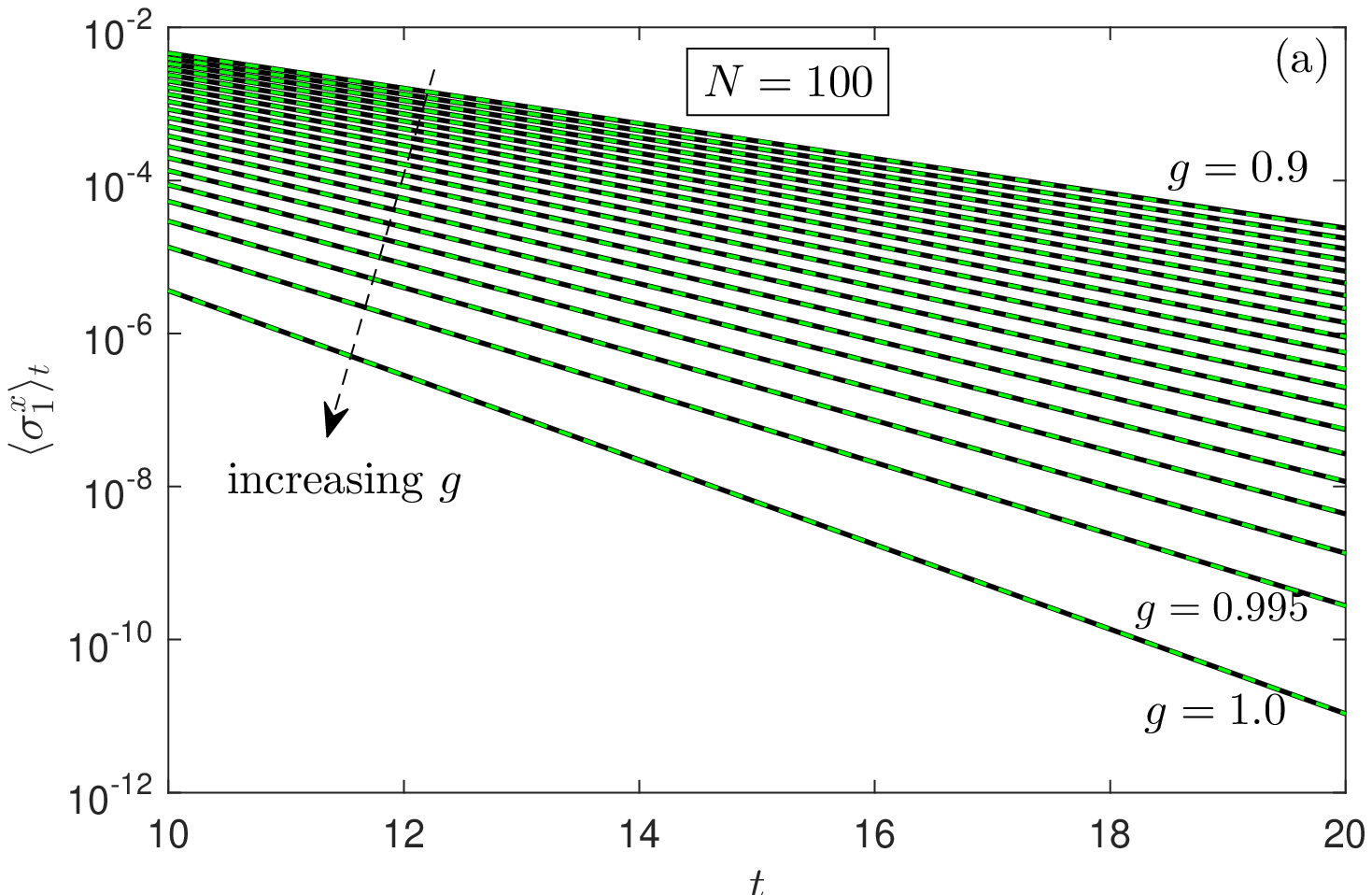}
\includegraphics[width=.53\textwidth]{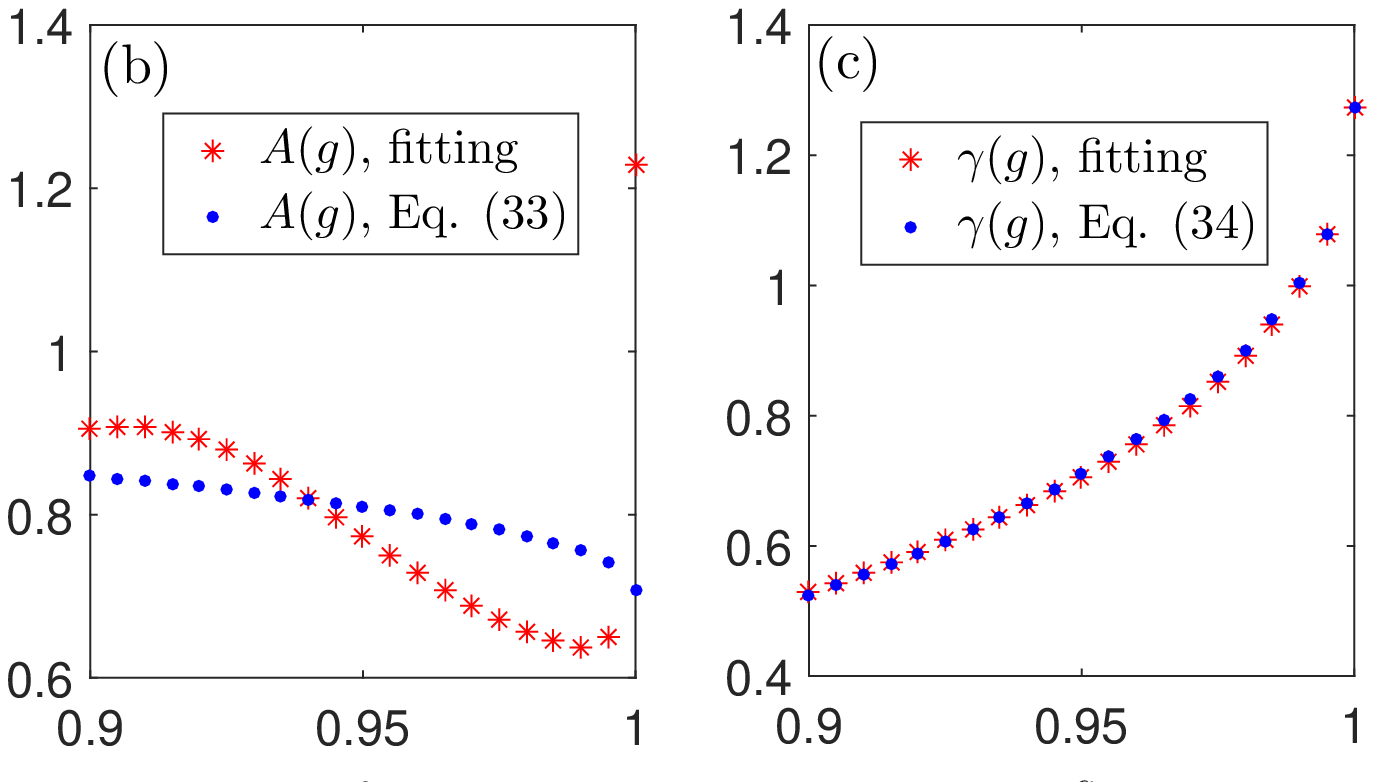}
\caption{(a) Long-time dynamics of $\langle\sigma^x_1\rangle_t$ after quenches into the vicinity of the critical point $g=1$ for a ring with $N=100$ sites. Results for $g=0.9,0.905,0.91,\cdots,1$ are shown. The dashed green curves show the corresponding exponential fitting. (b) The simulated prefactor $A(g)$ from the data fitting (red stars) and the one given by Eq.~(\ref{Ag}) (blue dots). (c) The simulated decay function $\gamma(g)$ from the data fitting (red stars) and the one given by Eq.~(\ref{rg}) (blue dots).}
\label{Fig7}
\end{figure}

 To check the validity of Eqs.~(\ref{Ag}) and (\ref{rg}), we calculate $\langle\sigma^x_1\rangle_t$ for a larger ring with $N=100$ sites within the time interval $t\in[10,20]$ (lies in the long-time regime) and perform exponential fittings of the data [Fig.~\ref{Fig7}(a)]. This results in the simulated $A(g)$ and $\gamma(g)$ presented in Figs.~\ref{Fig7}(b) and (c), respectively. In Fig.~\ref{Fig7}(b) we see the discontinuity revealed in Ref.~\cite{Rossini2020} in $A(g\to 1^-)$, as well as a discrepancy between the simulated result and that given by Eq.~(\ref{Ag}), which is believed to be caused by the finite-size nature of our simulations. Nevertheless, the decay function $\gamma(g)$ is found to agree well with Eq.~(\ref{rg}) [Fig.~\ref{Fig7}(c)]. We have checked that similar discontinuities in the prefactor and decay function also exist in the quenched dynamics of the string operator $X_j$ (data not shown). We believe such singular behaviors can be manifested in quench dynamics of generic odd operators that generally decay in the long-time limit.

\section{Reduced dynamics of two nearest-neighbor spins}\label{SecIV}
\subsection{Determination of the two-site reduced density matrix}
\par In this section, we study the dynamics of the reduced density matrix $\rho^{(12)}(t)$ of two nearest-neighbor spins, say $\vec{\sigma}_1$ and $\vec{\sigma}_2$. In the standard basis $\{|\uparrow\rangle_1|\uparrow\rangle_2,|\uparrow\rangle_1|\downarrow\rangle_2,|\downarrow\rangle_1|\uparrow\rangle_2,|\downarrow\rangle_1|\downarrow\rangle_2\}$, the matrix elements of $\rho^{(12)}(t)$ can be expressed as time-dependent expectation values of products of suitable fermion operators~\cite{PRA2010}:
\begin{eqnarray}
\rho^{(12)}_{11}(t)&=&\langle \sigma^+_1\sigma^-_1\sigma^+_2\sigma^-_2\rangle_t=\langle c^\dag_1 c_1 c^\dag_2c_2\rangle_t,\nonumber\\
\rho^{(12)}_{22}(t)&=& \langle \sigma^+_1\sigma^-_1\sigma^-_2\sigma^+_2\rangle_t=\langle c^\dag_1 c_1 c_2c^\dag_2\rangle_t,\nonumber\\
\rho^{(12)}_{33}(t)&=&\langle \sigma^-_1\sigma^+_1\sigma^+_2\sigma^-_2\rangle_t=\langle c_1 c^\dag_1c^\dag_2 c_2\rangle_t,\nonumber\\
\rho^{(12)}_{12}(t)&=&\langle\sigma^+_1\sigma^-_1\sigma^-_2\rangle_t=-\langle c^\dag_1c_1c_2\rangle_t,\nonumber\\
\rho^{(12)}_{13}(t)&=&\langle\sigma^-_1\sigma^+_2\sigma^-_2\rangle_t=\langle c_1 c^\dag_2 c_2\rangle_t,\nonumber\\
\rho^{(12)}_{14}(t)&=&\langle \sigma^-_1\sigma^-_2\rangle_t=-\langle c_1c_2\rangle_t,\nonumber\\
\rho^{(12)}_{23}(t)&=&\langle \sigma^-_1\sigma^+_2\rangle_t=-\langle c_1c^\dag_2\rangle_t,\nonumber\\
\rho^{(12)}_{24}(t)&=&\langle \sigma^-_1\sigma^-_2\sigma^+_2\rangle_t=\langle c_1c_2c^\dag_2\rangle_t,\nonumber\\
\rho^{(12)}_{34}(t)&=&\langle \sigma^-_1\sigma^+_1\sigma^-_2\rangle_t=\langle c_1c^\dag_1 c_2\rangle_t.
\end{eqnarray}
The remaining matrix elements are determined by the Hermitian property of $\rho^{(12)}(t)$. From Eq.~(\ref{sym}) we have $\rho^{(12)}_{12}=\rho^{(12)}_{13}$, $\rho^{(12)}_{24}=\rho^{(12)}_{34}$, and $\rho^{(12)}_{22}=\rho^{(12)}_{33}$. At first sight, it seems difficult to evaluate the off-diagonal elements $\rho^{(12)}_{12}$ and $\rho^{(12)}_{24}$ since they involve three fermionic operators. Thanks to the translational invariance of the state, we can use $c^\dag_1c_1=(\sigma^z_1+1)/2$ and $\sigma^-_2=(-\sigma^z_1)c_2$ to rewrite $\rho^{(12)}_{12}$ as
\begin{eqnarray}\label{rho12}
\rho^{(12)}_{12}&=& -\frac{1}{2}\langle c_2\rangle_t+\frac{1}{2}\langle\sigma^-_2\rangle_t\nonumber\\
&=&-\frac{1}{2}\langle c_2\rangle_t+\frac{1}{2}\langle\sigma^-_1\rangle_t\nonumber\\
&=&\frac{1}{2}(\langle c_1\rangle_t-\langle c_2\rangle_t).
\end{eqnarray}
Similarly,
\begin{eqnarray}\label{rho24}
\rho^{(12)}_{24}&=& \langle c_1 \rangle_t-\langle c_1 c^\dag_2c_2\rangle_t\nonumber\\
&=&\langle c_1 \rangle_t- \rho^{(12)}_{13} \nonumber\\
&=&\frac{1}{2}(\langle c_1\rangle_t+\langle c_2\rangle_t).
\end{eqnarray}
We see that, due to the symmetries of the Hamiltonian and the initial state, the expectation values of triple-operators can indeed be expressed as those of single-operators, which can be directly calculated via Eq.~(\ref{lengthy}).
\par The off-diagonal elements $\rho^{(12)}_{14}$ and $\rho^{(12)}_{23}$ involves two fermionic operators and can be easily calculated through similar procedures as in obtaining $\langle\sigma^z_1\rangle_t$:
\begin{eqnarray}\label{rho1423}
\rho^{(12)}_{14}&=&\frac{1}{N} \sum_{\sigma=\pm}\sum_{k \in K'_\sigma} \sin k u^{(\sigma)*}_k v^{(\sigma)}_k,\nonumber\\
\rho^{(12)}_{23}&=&-\frac{1}{2N}\left(2\sum_{\sigma=\pm}\sum_{k \in K'_\sigma} \cos k |u^{(\sigma)}_k |^2-1\right).
\end{eqnarray}
Note that $\rho^{(12)}_{23}$ is indeed real.
\par The calculation of the diagonal element $\rho^{(12)}_{11}$, which involves the product of four fermion operators, is more complicated. After a tedious but straightforward calculation, we get (see Appendix~\ref{AppA} for some details)
\begin{eqnarray}\label{rho11}
&&\rho^{(12)}_{11}=\frac{4}{N^2}\sum_{\sigma=\pm} \sum_{k>k'\in K'_\sigma }(1-\cos k\cos k')|v^{(\sigma)}_{k}|^2|v^{(\sigma)}_{k'}|^2\nonumber\\
&&+\frac{4}{N^2}\sum_{\sigma=\pm} \sum_{k>k'\in K'_\sigma}\sin k\sin k' \Re[u^{(\sigma)*}_{k}v^{(\sigma)}_{k}u^{(\sigma)}_{k'} v^{(\sigma)*}_{k'}]\nonumber\\
&&+\frac{2}{N^2}\sum_{k\in K'_+ }\sin^2k |v^{(+) }_{k}|^2\nonumber\\
&&+\frac{2}{N^2}\sum_{k\in K'_-}(2+\cos k)(1-\cos k) |v^{(-) }_{k}|^2.
\end{eqnarray}
It is interesting to note that in the thermodynamic limit $\rho^{(12)}_{11}$ can be expressed in terms of double integrals over $k$ and $k'$ (using $\frac{4}{N^2}\sum_{k>k'\in K'_\sigma}\to\frac{1}{\pi^2}\int^\pi_0dk\int^k_0dk'$):
\begin{eqnarray}\label{rho11TL}
&&\lim_{N\to\infty}\rho^{(12)}_{11}(t)\nonumber\\
&=&\frac{2}{\pi^2}\int^\pi_0dk\int^k_0dk' (1-\cos k\cos k')|v_{k}|^2|v_{k'}|^2\nonumber\\
&&+\frac{2}{\pi^2}\int^\pi_0dk\int^k_0dk'\sin k\sin k' \Re(u^{ *}_{k}v^{ }_{k}u^{ }_{k'} v^{ *}_{k'}),
\end{eqnarray}
where we have dropped the last two terms in Eq.~(\ref{rho11}) since they are of order $O(1/N)$.
The element $\rho^{(12)}_{22}$ can be obtained as
\begin{eqnarray}\label{rho22}
\rho^{(12)}_{22}(t)&=& \langle c^\dag_1 c_1 \rangle_t-\langle c^\dag_1 c_1c^\dag_2 c_2\rangle_t\nonumber\\
&=&\frac{1}{2}+\langle\sigma^z_1 \rangle_t-\rho_{11}(t).
\end{eqnarray}
We thus fully determined the reduced density matrix $\rho^{(12)}(t)$:
\begin{eqnarray}\label{Rho12}
&&\rho^{(12)}(t)\nonumber\\
&=&\left(
                 \begin{array}{cccc}
                   \rho^{(12)}_{11} & \rho^{(12)}_{12} & \rho^{(12)}_{12} & \rho^{(12)}_{14} \\
                    \rho^{(12)*}_{12}  & \rho^{(12)}_{22}& \rho^{(12)}_{23} &  \rho^{(12)}_{24} \\
                   \rho^{(12)*}_{12} & \rho^{(12)}_{23} & \rho^{(12)}_{22} &  \rho^{(12)}_{24} \\
                   \rho^{(12)*}_{14}  &\rho^{(12)*}_{24} & \rho^{(12)*}_{24} &  1-\rho^{(12)}_{11}-2\rho^{(12)}_{22} \\
                 \end{array}
               \right),
\end{eqnarray}
with the nonvanishing entries $\rho^{(12)}_{12}$, $\rho^{(12)}_{24}$, $\rho^{(12)}_{14}$, $\rho^{(12)}_{23}$, $\rho^{(12)}_{11}$, and $\rho^{(12)}_{22}$ given by Eqs.~(\ref{rho12})-(\ref{rho22}).
\par Below we are interested in the dynamics of various two-site equal-time correlators:
\begin{eqnarray}
\label{CFzz}
\langle\sigma^z_1\sigma^z_2\rangle_t&=&4\rho^{(12)}_{11}(t)-2\langle\sigma^z_1\rangle_t-1,\\
\label{CFxx}
\langle\sigma^x_1\sigma^x_2\rangle_t&=&2[\Re\rho^{(12)}_{14}(t)+\rho^{(12)}_{23}(t)],\\
\label{CFxy}
\langle\sigma^x_1\sigma^y_2\rangle_t&=&-2\Im\rho^{(12)}_{14},\\
\label{CFxz}
\langle\sigma^x_1\sigma^z_2\rangle_t&=&\langle X_2\rangle_t=-2\Re\langle c_2\rangle_t,
\end{eqnarray}
and the pairwise entanglement measured by the concurrence~\cite{Wooters},
\begin{eqnarray}\label{conc}
C(t)=\max\{0,\sqrt{\lambda_1}-\sqrt{\lambda_2}-\sqrt{\lambda_3}-\sqrt{\lambda_4}\},
\end{eqnarray}
with $\lambda_1,\cdots,\lambda_4$ being the eigenvalues of the matrix $\rho^{(12)}(\sigma_y\otimes\sigma_y)\rho^{(12)*}(\sigma_y\otimes\sigma_y)$ arranged in descending order. In passing we mention that $\langle\sigma^z_j\rangle_t$, $\langle\sigma^x_j\sigma^x_{j+1}\rangle_t$, and $\langle\sigma^x_j\sigma^y_{j+1}\rangle_t$ have been recently obtained in the quantum Ising ring for general translationally invariant product initial states by solving a closed hierarchy of Heisenberg equations for operators forming the Onsager algebra~\cite{Oleg}.
\begin{figure}
\includegraphics[width=.49\textwidth]{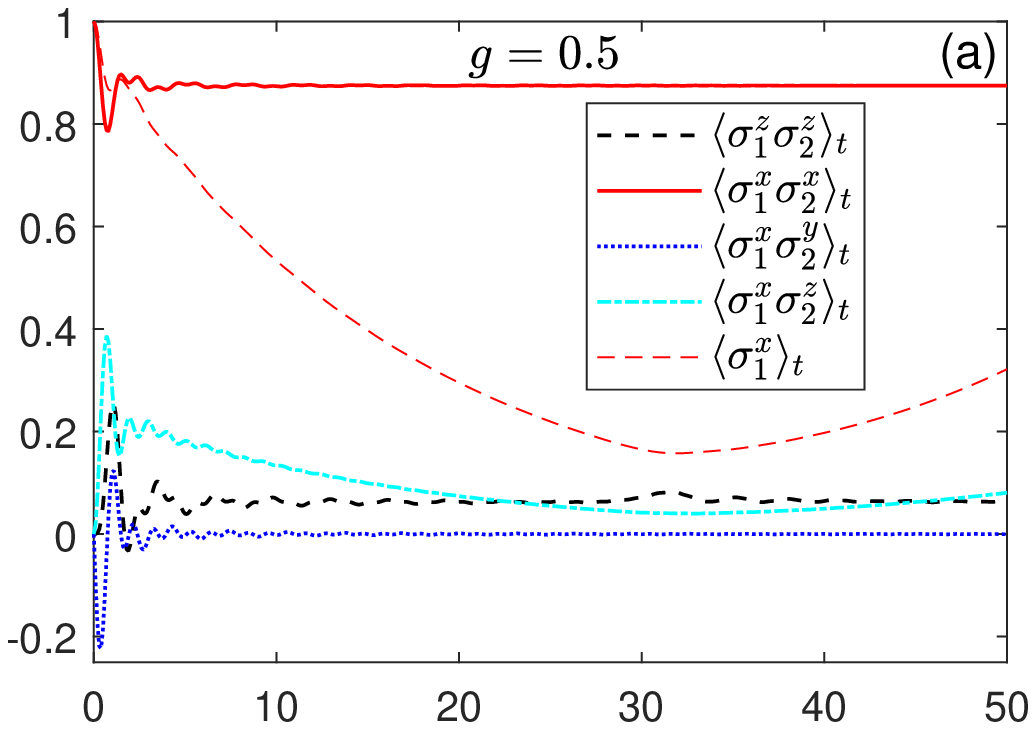}
\includegraphics[width=.49\textwidth]{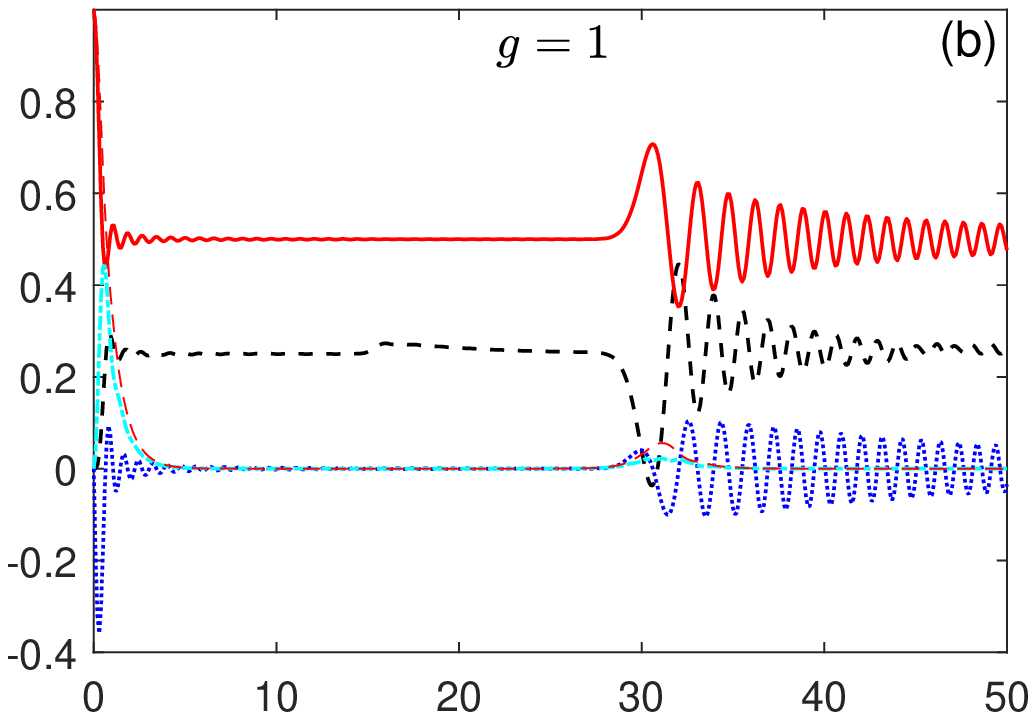}
\includegraphics[width=.49\textwidth]{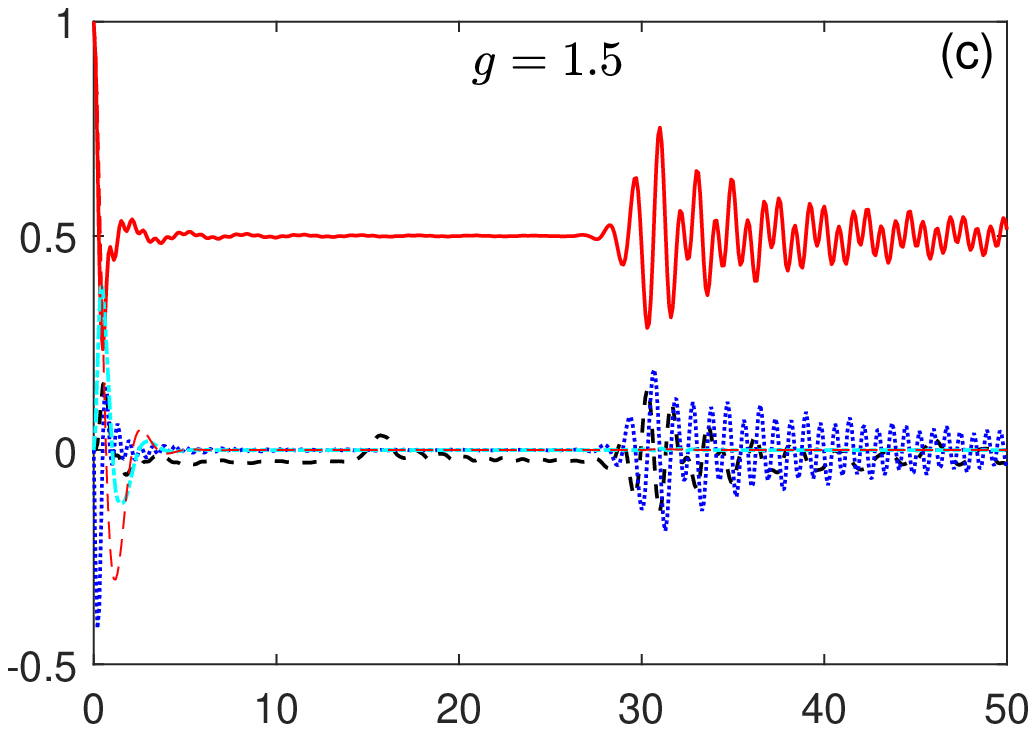}
\caption{Dynamics of equal-time correlators $\langle\sigma^z_1\sigma^z_2\rangle_t$, $\langle\sigma^x_1\sigma^x_2\rangle_t$, $\langle\sigma^x_1\sigma^y_2\rangle_t$, and $\langle\sigma^x_1\sigma^z_2\rangle_t$ after quenches to (a) $g=0.5$, (b) $g=1$, (c) $g=1.5$. The longitudinal magnetization dynamics $\langle\sigma^x_1\rangle_t$ is also presented. Numerical simulations are performed for $N=60$.}
\label{Fig8}
\end{figure}
\begin{figure}
\includegraphics[width=.52\textwidth]{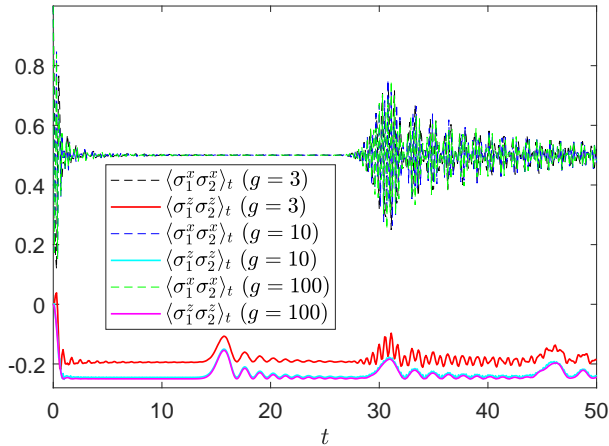}
\caption{Dynamics of $\langle\sigma^x_1\sigma^x_2\rangle_t$ and $\langle\sigma^z_1\sigma^z_2\rangle_t$ after quenches to the disordered phase with $g=3$, $10$, and $100$. Numerical simulations are performed for $N=60$.}
\label{Fig9}
\end{figure}
\subsection{Numerical results}
\begin{figure}
\includegraphics[width=.52\textwidth]{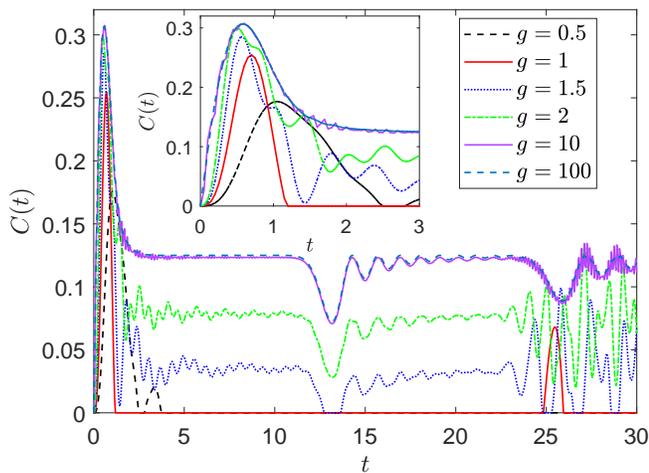}
\caption{Pairwise concurrence as a function of time after quenches to various values of $g$. The inset shows the initial evolution up to $t=3$. Numerical simulations are performed for $N=50$.}
\label{Fig10}
\end{figure}
\par Figure~\ref{Fig8}(a) shows the time dependence of $\langle\sigma^z_1\sigma^z_2\rangle_t$, $\langle\sigma^x_1\sigma^x_2\rangle_t$, $\langle\sigma^x_1\sigma^y_2\rangle_t$, and $\langle\sigma^x_1\sigma^z_2\rangle_t$ after a quench to $g=0.5$. The longitudinal magnetization $\langle\sigma^x_1\rangle_t$ is also presented for comparison. Although the longitudinal magnetization $\langle\sigma^x_1\rangle_t$ decays and exhibits a nonmonotonic behavior due to the finite-size effect, the equal-time correlator $\langle\sigma^x_1\sigma^x_2\rangle_t$ persists at a steady value $\sim0.875$. As the average of an odd operator, $\langle\sigma^x_1\sigma^z_2\rangle_t$ experiences an initial oscillatory behavior followed by an exponential decay. The transverse correlator $\langle\sigma^z_1\sigma^z_2\rangle_t$ is established with positive but small values after an initial oscillation.
\par For quenches to the critical point $g=1$ and to the disordered phase with $g=1.5$ [Figs.~\ref{Fig8}(b) and (c)], $\langle\sigma^x_1\sigma^x_2\rangle_t$ is still robust ($\sim0.5$) before the revival features start after $t\approx N/2$, even though the corresponding $\langle\sigma^x_1\rangle_t$ is vanishingly small. Figure~\ref{Fig9} shows the time evolution of $\langle\sigma^x_1\sigma^x_2\rangle_t$ and $\langle\sigma^z_1\sigma^z_2\rangle_t$ for quenches to large values of $g$. We see that $\langle\sigma^x_1\sigma^x_2\rangle_t$ exhibit a nearly perfect collapse-revival behavior and tends to converge for large $g$. It is expected that $\langle\sigma^x_1\sigma^x_2\rangle_t$ will acquire a steady value $\sim 0.5$ in the thermodynamic limit. Actually, from Eqs.~(\ref{rho1423}) and (\ref{CFxx}) we get
\begin{eqnarray}\label{CFxxTL}
\lim_{N\to\infty}\langle\sigma^x_1\sigma^x_2\rangle_t=1-\frac{4g^2}{\pi}\int^\pi_0dk\sin^2k\frac{1-\cos 2\Lambda_kt}{\Lambda^2_k}.
\end{eqnarray}
For $g\gg1$ and in the long-time limit $t\to\infty$, the term $\cos 2\Lambda_kt$ is a fast oscillating function of $k$, so that its integral tends to zero rapidly. Hence,
\begin{eqnarray}\label{CFxxTL1}
\lim_{N\to\infty}\langle\sigma^x_1\sigma^x_2\rangle_{t\to\infty}\approx 1-\frac{4g^2}{\pi}\int^\pi_0dk\sin^2k\frac{1 }{4g^2}=\frac{1}{2}.
\end{eqnarray}

\par The overall profile of $\langle\sigma^z_1\sigma^z_2\rangle_t$ also moves down as $g$ increases and tends to be saturated for large $g$. It is interesting to note that the saturated value of $\langle\sigma^z_1\sigma^z_2\rangle_t$ for large $g$ is about $-0.25$, indicating that some `antiferromagnetic order' is established quickly after the quench. Thus, although both $\langle\sigma^x_1\rangle_t$ and $\langle\sigma^z_1\rangle_t$ approaches zero in the large $g$ limit, their nearest-neighbor equal-time correlation functions are quite robust.
\par We now turn to discuss the entanglement generation after the quantum quench. In Fig.~\ref{Fig10} we show the dynamics of the nearest-neighbor concurrence $C(t)$ after quenches to various values of $g$, ranging from weak to strong fields. For quenches within the ordered phase, $C(t)$ reaches its maximum short after the quench starts and disappears for a long period of time. The entanglement dynamics shows similar short-time behavior for $g=1$, but exhibits a partial revival at $t\approx N/2$. However, for $g>1$ finite amount of entanglement tends to be generated after $C(t)$ passes its first peak, with a finite plateau value that increases with increasing $g$. The plateau value is expected to be $\sim0.125$ (expect for a dip around $t=N/4$) in the limit $g\to\infty$. The inset of Fig.~\ref{Fig10} shows the short-time window covering the first maxima of $C(t)$. It is observed that for a larger $g$ the first maximum of $C(t)$ appears earlier and has a higher value. Since it is experimentally simple to prepare our initial state, the sudden quench in the transverse field provides a possible way to generate long-lasting steady pairwise entanglement.
\section{Conclusions and Discussions}\label{SecV}
\par In this work, we obtained exact reduced dynamics of a single-spin and of two nearest-neighbor spins in a quantum Ising ring after a sudden quench of the transverse field. The quench starts with the ground state of the classical Ising ring (no field) and ends up with a finite value of the field. The initial state is chosen as a $Z_2$ symmetry-breaking fully ordered state. By writing the initial state as an equally weighted linear superposition of the two momentum-space ground states with distinct fermion parity, we analytically obtained the time-evolved state. Based on this, we derive analytical expressions for the time-dependent expectation values of all the relevant even operators. The dynamics of the relevant odd operators is obtained in a semi-analytical way using a recently developed Pfaffian method.
\par Having the obtained single-spin and two-spin reduced density matrices in hand, we thoroughly investigate quench dynamics of the magnetizations, single-spin purity, a string operator $X_j$ of length $j$, two-site equal-time correlator, and nearest-neighbor entanglement after quenches to various values of the field. Expectation values of generic odd operators are found to decay exponentially to zero in the long-time limit, consistent with observations in previous literature~\cite{JSM2012,Rossini2020}. The single-spin purity dynamics is determined by different components of the polarization for quenches to different phases. For quenches to large enough values of the field, the single-spin state tends to be maximally mixed; and the transverse and longitudinal two-spin correlators saturate to finite values in the thermodynamic limit. These asymptotic behaviors are quantitatively interpreted using the corresponding analytical expressions in the long-time limit. We also calculated the nearest-neighbor entanglement dynamics and find that quenching to the disordered phase provides a useful protocol to generate finite amount of entanglement over long periods of time.
\par Special attention is paid to quenches into the vicinity of the critical point. By performing large-scale simulations of the long-time quench dynamics, we quantitatively confirm an asymptotic exponential decay of the longitudinal magnetization derived in Ref.~\cite{JSM2012} in the thermodynamic limit. The decay function is found to agree well with the analytical results, though discrepancy in the prefactor is observed due to the finite-size effect. Nevertheless, we still confirm a recently discovered discontinuity in the prefactor~\cite{Rossini2020} for a quench to exactly the critical point. We study in detail the dynamical behaviors of the string operator $X_j$. The first maximum of $\langle X_j\rangle_t$ after a quench to the critical point is found to exponentially decrease with increasing string length, while the positions of these maxima increases linearly with $j$. The long-time dynamics of $\langle X_j\rangle_t$ exhibit a $j$-independent exponential decay with a smaller prefactor than the longitudinal magnetization. We also observe intermediate collapse of $\langle X_j\rangle_t$ followed by a partial revival at times which are multiples of $N/2$. These behaviors are consistent with a conformal field theory analysis by Cardy~\cite{Cardy2014}.
\par We conclude with some possible applications of our method. Recently, the generation of multipartite entanglement in integrable systems after a quantum quench has attracted growing attention~\cite{JSM2017}. As a measure for multipartite entanglement, it is useful to calculate the dynamics of the quantum Fisher information~\cite{QIF} after a quench. This involves of calculation of matrix elements of some observable between eigenstats of certain density matrix. Our approach may provide an efficient way for the calculation of these matrix elements for odd operators. As demonstrated by our study of the string operator $X_j$, our method is also applicable in the study of thermalization and relaxation to a steady state in spin chain systems, where efficient evaluation of long-time dynamics of local operators in large systems is desirable.

\section*{Acknowledgements}
We thank O. Lychkovskiy and D. Rossini for useful discussions. This work was supported by the Natural Science Foundation of China (NSFC) under Grant No. 11705007, and partially by the Beijing Institute of Technology Research Fund Program for Young Scholars.
\appendix
\begin{widetext}
\section{Calculation of $\rho^{(12)}_{11}(t)$}\label{AppA}
To derive Eq.~(\ref{rho11}), we perform the Fourier transforms of the real-space fermion operators to get
\begin{eqnarray}\label{rho11app}
\rho^{(12)}_{11}&=&-\langle c^\dag_1 c^\dag_2 c_1 c_2\rangle_t\nonumber\\
&=& -\frac{1}{2}\frac{1}{N^2}\sum_{k_1k_2k_3k_4\in K_+}e^{-ik_1-i2k_2+ik_3+i2k_4}\langle \phi_+| c^\dag_{k_1+}c^\dag_{k_2+} c_{k_3+}c_{k_4+} |\phi_+\rangle\nonumber\\
&&-\frac{1}{2}\frac{1}{N^2}\sum_{k_1k_2k_3k_4\in K_-}e^{-ik_1-i2k_2+ik_3+i2k_4}\langle \phi_-| c^\dag_{k_1-}c^\dag_{k_2-} c_{k_3-}c_{k_4-} |\phi_-\rangle.
\end{eqnarray}
Let us focus on the first term on the right side of the above equation. From $|\phi_+\rangle=\prod_{k\in K'_+}(u^{(+)}_+|\mathrm{vac}\rangle_{k+}+v^{(+)}_k|k,-k\rangle_+)$, we have to distinguish two cases:
\par 1) If $k_3\neq-k_4$, then $k_3$ and $k_4$ belong to different mode pairs. Hence, we must have $(k_1,k_2)=(k_3,k_4)$ or $(k_1,k_2)=(k_4,k_3)$ in order to get nonvanishing contributions:
\begin{eqnarray}
&&\sum_{k_1k_2k_3k_4\in K_+, k_3\neq-k_4}e^{-ik_1-i2k_2+ik_3+i2k_4}\langle \phi_+| c^\dag_{k_1+}c^\dag_{k_2+} c_{k_3+}c_{k_4+} |\phi_+\rangle\nonumber\\
&=&\sum_{ k_3k_4\in K_+, k_3\neq-k_4}(e^{-ik_3-i2k_4+ik_3+i2k_4}\langle \phi_+| c^\dag_{k_3+}c^\dag_{k_4+} c_{k_3+}c_{k_4+} |\phi_+\rangle+e^{-ik_4-i2k_3+ik_3+i2k_4}\langle \phi_+| c^\dag_{k_4+}c^\dag_{k_3+} c_{k_3+}c_{k_4+} |\phi_+\rangle)\nonumber\\
&=&-\sum_{ k_3k_4\in K_+, k_3\neq-k_4} [1-e^{-i(k_3-k_4)}] \langle \phi_+| c^\dag_{k_3+} c_{k_3+}c^\dag_{k_4+}c_{k_4+} |\phi_+\rangle \nonumber\\
&=&-\sum_{k_3>0,k_4>0,k_3\neq k_4} [1-e^{-i(k_3-k_4)}] |v^{(+)}_{k_3}|^2|v^{(+)}_{k_4}|^2- \sum_{k_3>0,k_4<0,k_3\neq -k_4} [1-e^{-i(k_3-k_4)}] |v^{(+)}_{k_3}|^2|v^{(+)}_{-k_4}|^2\nonumber\\
& &-\sum_{k_3<0,k_4>0,k_3\neq -k_4} [1-e^{-i(k_3-k_4)}] |v^{(+)}_{-k_3}|^2|v^{(+)}_{k_4}|^2- \sum_{k_3<0,k_4<0,k_3\neq k_4} [1-e^{-i(k_3-k_4)}] |v^{(+)}_{-k_3}|^2|v^{(+)}_{-k_4}|^2\nonumber\\
&=&-\sum_{k_3>0,k_4>0,k_3\neq k_4} \{[1-e^{-i(k_3-k_4)}]+[1-e^{-i(k_3+k_4)}]+[1-e^{-i(-k_3+k_4)}]+[1-e^{ i(k_3-k_4)}]\}|v^{(+)}_{k_3}|^2|v^{(+)}_{k_4}|^2 \nonumber\\
&=&-2\sum_{k_3>0,k_4>0,k_3\neq k_4} [1-\cos(k_4+k_3)+1-\cos(k_4-k_3)]|v^{(+)}_{k_3}|^2|v^{(+)}_{k_4}|^2 \nonumber\\
&=&-8\sum_{k>k'\in K'_+} (1-\cos k\cos k')|v^{(+)}_{k}|^2|v^{(+)}_{k'}|^2.
\end{eqnarray}
\par 2) If $k_3=-k_4$, we must have $k_1=-k_2$:
\begin{eqnarray}
&&\sum_{k_1k_2k_3k_4\in K_+, k_3=-k_4}e^{-ik_1-i2k_2+ik_3+i2k_4}\langle \phi_+| c^\dag_{k_1+}c^\dag_{k_2+} c_{k_3+}c_{k_4+} |\phi_+\rangle\nonumber\\
&=&\sum_{ k_2k_4 \in K_+ }e^{ik_2-i2k_2-ik_4+i2k_4}\langle \phi_+| c^\dag_{-k_2,+}c^\dag_{k_2+} c_{-k_4,+}c_{k_4+} |\phi_+\rangle\nonumber\\
&=&\sum_{k_2>0,k_4>0,k_2\neq k_4}  e^{-i(k_2-k_4)}v^{(+)*}_{k_2}(-u^{(+)}_{k_2})u^{(+)*}_{k_4}v^{(+)}_{k_4}-\sum_{k_4>0}|v^{(+)}_{k_4}|^2\nonumber\\
&&+\sum_{k_2<0,k_4>0,k_2\neq- k_4}  e^{-i(k_2-k_4)}v^{(+)*}_{-k_2}u^{(+)}_{-k_2}u^{(+)*}_{k_4}v^{(+)}_{k_4}+\sum_{k_4>0}e^{2ik_4}|v^{(+)}_{k_4}|^2\nonumber\\
&&+\sum_{k_2>0,k_4<0,k_2\neq- k_4}  e^{-i(k_2-k_4)}v^{(+)*}_{k_2}(-u^{(+)}_{k_2})u^{(+)*}_{-k_4}(-v^{(+)}_{-k_4})+\sum_{k_2>0}e^{-2ik_2}|v^{(+)}_{k_4}|^2\nonumber\\
&&+\sum_{k_2<0,k_4<0,k_2\neq k_4}  e^{-i(k_2-k_4)}v^{(+)*}_{-k_2}u^{(+)}_{-k_2}u^{(+)*}_{-k_4}(-v^{(+)}_{-k_4})-\sum_{k_4<0} |v^{(+)}_{-k_4}|^2\nonumber\\
&=&-8\sum_{k>k'\in K'_+}\sin k\sin k'\Re[u^{(+)*}_{k}v^{(+)}_{k}u^{(+)}_{k'} v^{(+)*}_{k'}]-4\sum_{k\in K'_+ }\sin^2k |v^{(+) }_{k}|^2.
\end{eqnarray}
As to the second term in Eq.~(\ref{rho11app}), we need to further take into account the contribution from $k_3=0$ or $k_4=0$ in the set $K_-$. By combining all these contributions, we finally obtain the result presented in Eq.~(\ref{rho11}).
\end{widetext}

\end{document}